\pdfoutput=1
\documentclass[fleqn,usenatbib]{mnras}

\usepackage{newtxtext,newtxmath}

\usepackage[T1]{fontenc}
\usepackage{ae,aecompl}

\usepackage{graphicx}	
\usepackage{amsmath}	
\usepackage{amssymb}	
\usepackage{longtable}
\usepackage[usenames, dvipsnames]{color}
\usepackage{times}
\usepackage{subfigure}
\usepackage{epsfig}
\usepackage{graphics}
\usepackage[toc,page]{appendix} 
\usepackage{hyperref}

\renewcommand{\descriptionlabel}[1]%
  {\hspace{\labelsep}\textbf{#1}}

\title[M13]{A new study of the variable star population in the Hercules globular cluster (M13; NGC 6205)$\thanks{Based on observations made with the telescope IAC80, operated by the Instituto de Astrof\'isica de Canarias in the Spanish
Observatorio del Teide on the island of Tenerife, and with the 2.0 m telescope at the Indian Astrophysical Observatory,
Hanle, India.}$}

\author[D. Deras et al.]{
D. Deras,$^{1}$\thanks{E-mail: dderas@astro.unam.mx, 
armando@astro.unam.mx, clh@iac.es, ivanbf@oac.unc.edu.ar,
calderon@oac.unc.edu.ar, 
smuneer@iiap.res.in, giridhar@iiap.res.in,}
A. Arellano Ferro$^{1}$, C. L\'azaro$^{2,3}$, I.H. Bustos Fierro$^{4}$, J. H. Calder\'on$^{4,5}$, \and 
S. Muneer$^{6}$ 
and Sunetra Giridhar$^{6}$\\
$^{1}$Instituto de Astronom\'ia, Universidad Nacional Aut\'onoma de M\'exico, Ciudad Universitaria, C.P. 04510, M\'exico.\\	
$^{2}$Departamento de Astrof\'isica, Universidad de La Laguna, E-38206 La Laguna, Tenerife, Spain.\\
$^{3}$Instituto de Astrof\'isica de Canarias (IAC), E-38205 La Laguna, Tenerife,
Spain.\\
$^{4}$Observatorio Astron\'omico, Universidad Nacional
de C\'ordoba, C\'ordoba, Argentina.\\
$^{5}$Consejo Nacional de Investigaciones Cient\'ificas y
T\'ecnicas (CONICET), C\'ordoba, Argentina.\\
$^{6}$Indian Institute of Astrophysics, Koramangala 560034, Bangalore, India.\\
}

\date{Accepted XXX. Received YYY; in original form ZZZ}

\pubyear{2018}
\begin{document}
\label{firstpage}
\pagerange{\pageref{firstpage}--\pageref{lastpage}}
	\maketitle

\begin{abstract}
We present the results from $VI$ CCD time-series photometry of the globular cluster M13 (NGC 6205). From the Fourier decomposition of the light curves of RRab and RRc stars we found an average metallicity of [Fe/H]zw = -1.58 $\pm$ 0.09. The distance to the cluster was estimated as 7.1 $\pm$ 0.1 kpc from independent methods related to the variable star families RR Lyrae, SX Phe and W Virginis, from the luminosity of the theoretical ZAHB and from the orbit solution of a newly discovered contact binary star. The RR Lyrae pulsation modes are segregated by the red edge of the first overtone instability strip in this OoII type cluster. A membership analysis of 52,800 stars in the field of the cluster is presented based on \textit{Gaia}-DR2 proper motions which enabled the recognition of 23,070 likely cluster members, for 7,630 of which we possess $VI$ photometry.  The identification of member stars allowed the construction of a clean CMD and a proper ZAHB and isochrone fitting, consistent with a reddening, age and distance of 0.02 mag, 12.6 Gyrs and 7.1 kpc respectively. We report seven new variables; one RRc, two SX Phe stars, three SR and one contact binary. V31 presents double-mode nature and we confirm V36 as RRd. Fifteen variable star candidates are also reported. The analysis of eighteen stars in the field of the cluster, reported as RR Lyrae from the $Gaia$-DR2 data base reveals that at least seven are not variable. We noted the presence of a high velocity star in the field of the cluster.
\end{abstract}




\section{Introduction}

M13 (NGC 6205) is a very bright globular cluster ($V\approx5.8$ mag) in the constellation of Hercules ($\alpha = 16^{h} 41' 41.24'', \delta = +36^{\circ} 27' 35.5''$, J2000), and located in the halo ($l = 59.01^{\circ}$, $b = 40.91^{\circ}$) of the Galaxy. 

The variable star population of M13 is not particularly rich, it contains RR Lyrae stars (9), SX Phe (4), CW stars (3), variable red giants (16) and pulsars (5) that have been reported in the literature. The Catalogue of Variable Stars in Globular Clusters (CVSGC; \citet{Clement2001}; 2015 edition) lists 53 variable stars although only 45 have been confirmed as variables.
In the Catalog of Parameters for Milky Way Globular Clusters compiled by  \citet{Harris1996} (2010 edition), the distance from the Sun and the reddening for the cluster are given as 7.1 kpc and $E(B-V) = 0.02$ respectively.
One of the main characteristics of the Colour-Magnitude Diagram (CMD) of M13 is its well-known extreme blue horizontal branch. 

Previous recent CCD photometric studies of M13 have been successful in identifying new variables, studying individual cases and filtering out some non-variables from the variable stars designations,
(\citealt{Kopacki2003}; \citealt{Kopacki2005})
. With the aim of determining the mean metallicity and distance to the cluster from specific calibrations valid for different families of variable stars, such as Fourier light curve decomposition, luminosity of the horizontal branch and P-L relations, the analysis of an extensive new time-series of CCD images is undertaken in the present work. Given this new data, a systematic search for new variables by several approaches may help updating
the variability census of M13.  We aim to  discuss the membership of the stars in the field of M13, particularily of the variable star population,  using the recent proper motions and radial velocities available in the $Gaia$-DR2.

The paper is structured in the following way: In $\S$ 2, we describe our observations, the data reduction process and the transformation to the standard photometric system. In $\S$ 3, we discuss the population of variable stars present in M13, as well as the strategies used to find new variables and their utility in estimating stellar and mean cluster  physical parameters. In $\S$ 4, we explain our methodology to estimate the values of the physical parameters of the RR Lyrae stars. In $\S$ 5, we mention different values for the metallicity of the cluster found in the literature and compare them with our own results. In $\S$ 6, we determine the distance to M13 using 7 independent methods. In $\S$ 7, we discuss the strategy used to determine the membership of the stars in the cluster. In $\S$ 8 we discuss the overall properties of the CMD of M13 as well as the structure of its horizontal branch and in $\S$ 9, we summarise our conclusions.
Finally, in Appendix A we discuss the presence of a high velocity star in the field of the cluster.

\section{Data, observations and reductions}

The observations used for the present work were performed at two different locations. The first set of data was obtained using the 2 m telescope at the Indian Astronomical Observatory (IAO) in Hanle, India on 6 nights separated into two seasons. The first season spans the nights of June 7th-9th 2014, the second season spans the nights of August 3rd-5th 2014. The detector used was a SITe ST-002 thinned backside illuminated CCD of 2048$\times$2048 pixels with a scale of 0.296 arcsec/pix, translating to a field of view (FoV) of approximately 10.1$\times$10.1~arcmin$^2$. 
The second set of data was obtained with the 0.80 m telescope at the Instituto de Astrof\'isica de Canarias (IAC) during 12 nights divided in four seasons. The first season spans the nights of June 13th-15th 2016, the second season the nights of June 26th-29th 2016, the third season the nights of July 2nd-4th 2016 and the fourth season spans the nights of July 16th and 18th 2016. We used the CAMELOT camera with  2048x2048 pixels and 0.304 arcsec/pixel, with a 10.4$\times$10.4 arcmin$^2$ FoV, and a back illuminated detector CCD42-40 from E2V Technologies. 

Table \ref{tab:observations} summarizes the observation dates, exposure times and average seeing conditions.

\begin{table}
\caption{The distribution of observations of M13 for each filter.
Columns $N_{V}$ and $N_{I}$ give the number of images taken with the $V$ and $I$
filters respectively. Columns $t_{V}$ and $t_{I}$ provide the exposure time,
or range of exposure times employed during each night for each filter. The 
average seeing is listed in the last column.}
\centering
\begin{tabular}{lccccc}
\hline
Date          &  $N_{V}$ & $t_{V}$ (s) & $N_{I}$ &$t_{I}$ (s)&Avg. seeing (") \\
\hline

20140607 & 16  & 40       & 17  & 7      & 1.6 \\
20140608 & 24  & 40       & 24  & 7      & 1.7 \\
20140609 & 2   & 40       & 4   & 7      & 1.7 \\
20140803 & 16  & 30       & 18  & 7      & 1.6 \\
20140804 & 24  & 30       & 26  & 7      & 1.7 \\
20140805 & 20  & 30       & 22  & 7      & 1.6 \\
20160613 & 6   & 600      & 6   & 400    & 1.5 \\
20160614 & 96  & 100-300  & 94  & 30-200 & 1.7 \\
20160615 & 86  & 80-100   & 67  & 20-30  & 1.7 \\
20160626 & 25  & 100      & 26  & 50     & 2.6 \\
20160627 & 94  & 60-80    & 93  & 15-30  & 1.2 \\ 
20160628 & 94  & 70-80    & 94  & 20-30  & 1.7 \\
20160629 & 56  & 150-200  & 33  & 60-80  & 2.4 \\
20160702 & 79  & 80-100   & 80  & 15-20  & 1.3 \\
20160703 & 113 & 40-80    & 116 & 10-20  & 1.2 \\
20160704 & 61  & 80-100   & 58  & 20-40  & 1.8 \\
20160716 & 41  & 120      & 45  & 80-100 & 1.8 \\
20160718 & 100 & 100      & 99  & 20-30  & 1.7 \\

\hline
Total:   &  953  & -- & 922   & -- & --  \\
\hline
\end{tabular}
\label{tab:observations}
\end{table}

\subsection{Difference Image Analysis}

For the reduction of our data, we employed the software Difference Imaging Analysis (DIA) with its pipeline implementation DanDIA \footnote{DanDIA is built from the DanIDL library of IDL routines
available at \url{http://www.danidl.co.uk}} (\citealt{Bramich2008}; \citealt{Bramich2013}). With this, we were able to obtain high-precision photometry for all the point sources in the FoV of the two CCDs at the two locations. First, a reference image is created by DanDIA by stacking the best images in each filter and then it subtracts it from the rest of the images. The differential flux for each star is then determined by means of the PSF calculated by DanDIA from about 300-400 isolated stars in the FoV. Differential fluxes are converted into total fluxes.  
The total flux $f_{\mbox{\scriptsize tot}}(t)$ in ADU/s at each epoch $t$ can be estimated as:
\begin{equation}
f_{\mbox{\scriptsize tot}}(t) = f_{\mbox{\scriptsize ref}} +
\frac{f_{\mbox{\scriptsize diff}}(t)}{p(t)},
\label{eqn:totflux}
\end{equation}

\noindent
where $f_{\mbox{\scriptsize ref}}$ is the reference flux (ADU/s), $f_{\mbox{\scriptsize diff}}(t)$ is the differential flux (ADU/s) and
$p(t)$ is the photometric scale factor (the integral of the kernel solution). Conversion to instrumental magnitudes was achieved using:
\begin{equation}
m_{\mbox{\scriptsize ins}}(t) = 25.0 - 2.5 \log \left[ f_{\mbox{\scriptsize tot}}(t)
\right],
\label{eqn:mag}
\end{equation}

\noindent
where $m_{\mbox{\scriptsize ins}}(t)$ is the instrumental magnitude of the star at time $t$. The above procedure has been described in detail in \cite{Bramich2011}. 

\begin{figure} 
\includegraphics[width=8.0cm,height=5.0cm]{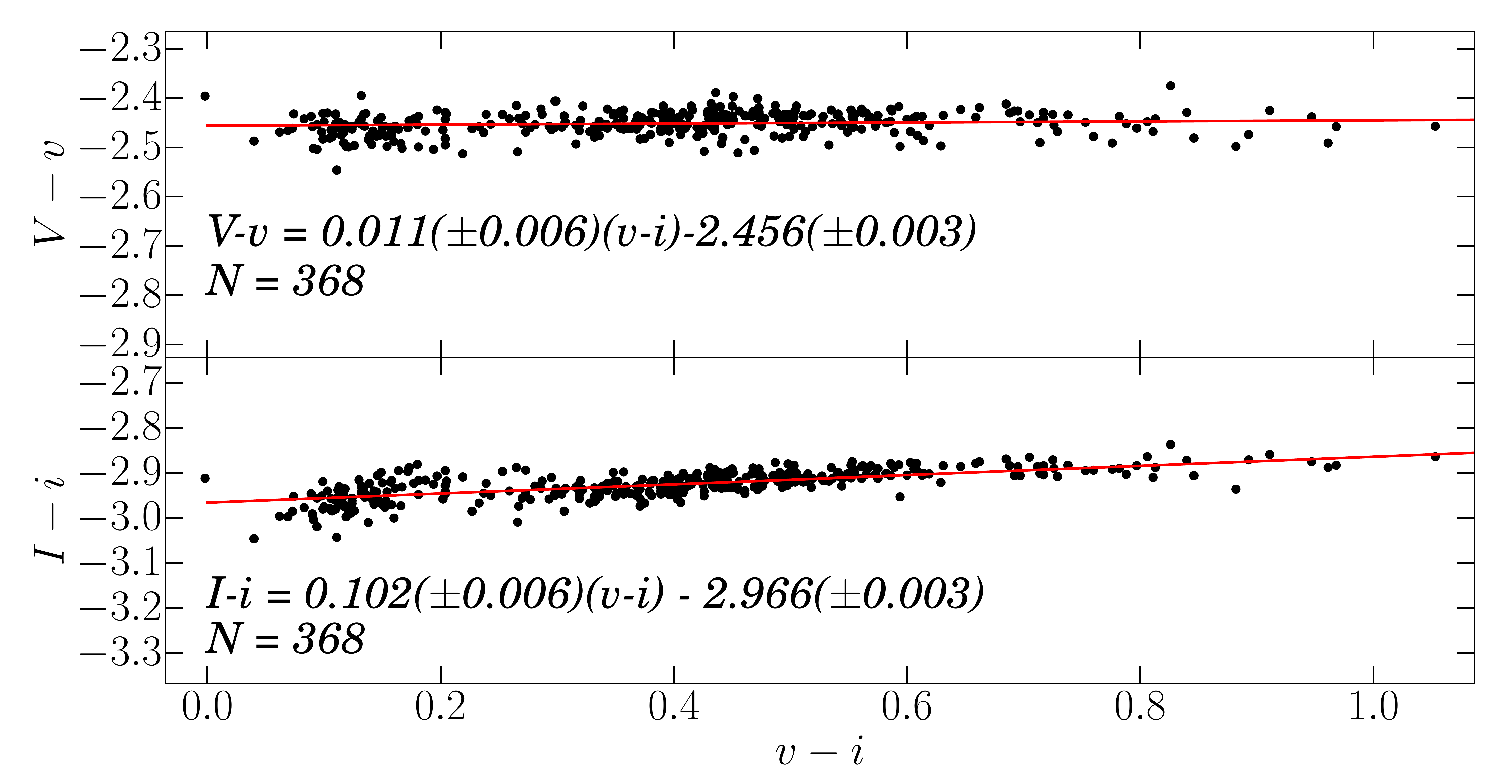}
\includegraphics[width=8.0cm,height=5.0cm]{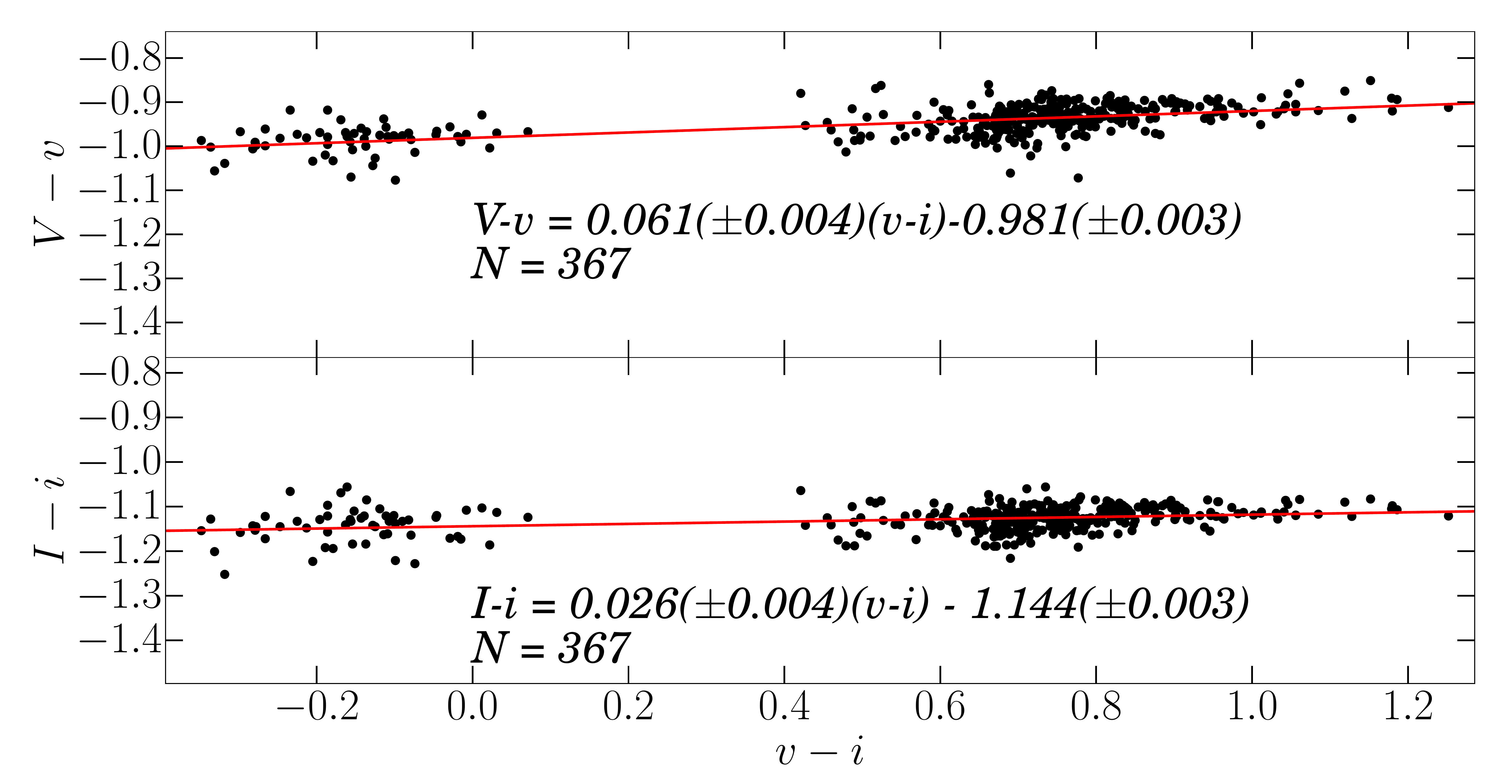}

\caption{Transformation relations obtained for the $V$ and $I$ filters between the instrumental and the standard photometric systems. To carry out the transformation, we made use of a set of standard stars (N = 368 for IAC and N = 367 for Hanle) in the field of M13. The top panel corresponds to the data from IAC and the bottom panel to the data from Hanle.}
    \label{transV}
\end{figure}
%

\subsection{Photometric Calibrations}

\subsubsection{Relative calibration}
\label{sec:rel}

To correct for possible systematic errors, we applied the methodology developed in \cite{Bramich2012} to solve for the magnitude offsets $Z_{k}$ that should be applied to each photometric measurement from the image $k$. In terms of DIA, this translates into a correction for the systematic error introduced into the photometry due to a possible error in the flux-magnitude conversion factor \citep{Bramich2015}. In the present case the corrections were very small, $\sim$ 0.001 mag for stars brighter than $\sim V =18.0$.

\subsubsection{Absolute calibration}
\label{absolute}

Standard stars in the field of M13 are included in the online collection of \cite{Stetson2000} \footnote{\url{http://www3.cadc-ccda.hia-iha.nrc-cnrc.gc.ca/community/STETSON/standards}} and we used them to transform instrumental $vi$ magnitudes into the Johnson-Kron-Cousins standard \emph{VI} system. The mild colour dependence of the standard minus instrumental magnitudes is shown in Fig. \ref{transV} for both the IAC and Hanle observations. The transformation equations are explicitly given in the figure itself.

\section{Variable stars in M13}

\begin{figure*} 
\includegraphics[width=18.0cm,height=9.5cm]{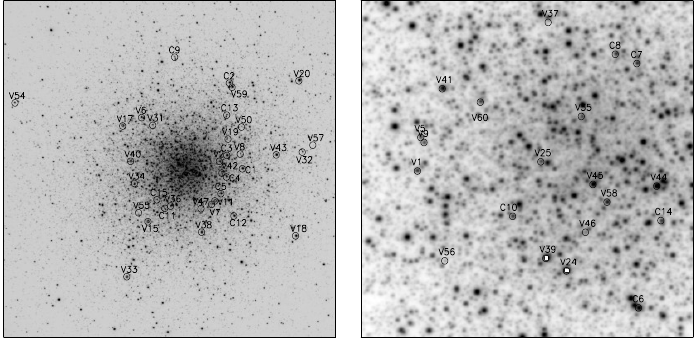}
\caption{Identification chart of all known variables in M13 listed in Table \ref{variables} and the candidate stars listed in Table \ref{candidatas}. The left panel is a field of 10.4$\times$10.4 arcmin$^2$. The right panel is 2.0$\times$2.0 arcmin$^2$.
North is up and East is to the left.}
\label{M13_CHART}. 
\end{figure*}

All the previously known and newly discovered variable stars in M13, are listed in Table \ref{variables} and have been identified in Fig. \ref{M13_CHART}.

\begin{table*}
\begin{center}
\caption{Data of variable stars in M13 in the FoV of our images.}
\label{variables}

\begin{tabular}{llccccccccc}
\hline
Star ID & Type & $<V>$ & $<I>$   & $A_V$  & $A_I$ & $P$    &  $\rm HJD_{\rm max}$ & $\alpha$ (J2000.0)  & $\delta$  (J2000.0)   &  \textit{Gaia}-DR2 Source    \\
        &      & (mag) & (mag)   & (mag)  & (mag) & (days) &  + 2450000 &                     &         \\
  
 \hline
V1           & CW            & 14.04     & 13.49 & 1.04 & 0.71 & 1.4590      & 7573.7624     & 16:41:46.47 & 36:27:27.59 & 1328057184175359488\\
V2           & CW            & 12.99     & 12.32 & 0.92 & 0.66 & 5.1108      & 7555.5896     & 16:41:35.89 & 36:27:48.27 & 1328057867076964864\\
V5           & RRc \emph{Bl} & 14.79     & 14.37 & 0.53 & 0.34 & 0.381784    & 7555.5102     & 16:41:46.36 & 36:27:39.76 & 1328057179882276480\\
V6           & CW            & 14.08$^1$ & 13.41 &   -- &   -- & 2.1129      & 7554.6234     & 16:41:47.96 & 36:29:09.50 & 1328059413265281920\\
V7           & RRc           & 14.90     & 14.53 & 0.34 & 0.26 & 0.312668    & 7569.5321     & 16:41:37.11 & 36:26:28.71 & 1328057768297931904\\
V8           & RRab          & 14.84     & 14.25 & 0.85 & 0.55 & 0.750303    & 7568.5076     & 16:41:32.64 & 36:28:01.92 & 1328058077530125568\\ 
V9           & RRc \emph{Bl} & 14.82     & 14.37 & 0.53 & 0.36 & 0.392724    & 7566.5022     & 16:41:46.25 & 36:27:37.75 & 1328057184175404544\\   
V11$^{1}$    & SR            & 11.86     & 10.34 &>0.06 &>0.05 & 92.0$^{4}$  &--             & 16:41:36.63 & 36:26:35.51 & 1328057763997754112\\
V15$^{1}$    & SR            & 12.09     & 10.67 &>0.01 &>0.01 & 30.0$^{4}$  &--             & 16:41:47.02 & 36:25:57.22 & 1328057081103004032\\
V17$^{1}$    & SR            & 11.93     & 10.46 &>0.05 &>0.02 & 43.0$^{4}$  &--             & 16:41:50.91 & 36:28:54.18 & 1328058696010941568\\
V18          & L             & 12.29     & 10.95 &>0.12 &>0.05 & 41.25$^{4}$ &--             & 16:41:24.07 & 36:25:30.52 & 1328057523479525760\\
V19$^{1}$    & SR            & 12.00     & 10.53 &>0.02 &>0.002& 30.0        &--             & 16:41:31.99 & 36:28:29.82 & 1328058077536075264\\
V20          & SR            & 12.06     & 10.54 &>0.11 &>0.06 & 40.0$^{4}$  &--             & 16:41:23.52 & 36:30:17.04 & 1328105150377778048\\
V24$^{1}$    & SR            & 11.94     & 10.36 &>0.05 &>0.03 & 45.0$^{4}$  &--             & 16:41:41.94 & 36:26:51.76 & 1328057149822883328\\
V25          & RRc           & 14.62     & 14.23 & 0.50 & 0.30 & 0.429529    & 7566.5022     & 16:41:42.70 & 36:27:30.86 & 1328057905736994560\\ 
V31          & RRd           & 14.39     & 13.81 & 0.034& 0.018& 0.32904     & 7567.4298     & 16:41:46.26 & 36:28:55.11 & --                 \\  
             &               &           &       & 0.048& 0.020& 0.31936     &               &             &             &                    \\  
V32          & SR            & 14.14     & 13.26 &>0.03 &>0.02 & 33.0        &--             & 16:41:23.03 & 36:28:05.10 & 1328104669341378432\\
V33$^{1}$    & SR            & 11.98     & 10.49 &>0.11 &>0.07 & 33.0        &--             & 16:41:50.26 & 36:24:15.37 & 1328056703145773696\\
V34          & RRc \emph{Bl} & 14.81     & 14.34 & 0.40 & 0.32 & 0.389497    & 7568.4786     & 16:41:49.08 & 36:27:07.58 & 1328057111162818560\\  
V35          & RRc           & 14.81     & 14.54 & 0.22 & 0.18 & 0.319991    & 7566.4249     & 16:41:41.45 & 36:27:47.15 & 1328057905737064576\\  
V36          & RRd           & 14.82     & 14.51 & 0.047 & 0.034 & 0.31596   & 7566.4299     & 16:41:43.48 & 36:26:25.97 & 1328057149822551424\\  
             &               &           &       & 0.058 & 0.032 & 0.30424   &               &             &             &                    \\  
V37          & SX Phe        & 17.18     & --    & 0.07 & 0.06 & 0.049411    &--             & 16:41:42.45 & 36:28:21.50 & --                 \\ 
V38          & SR            & 12.13     & 10.66 &>0.16 &>0.12 & 32.0$^{4}$  &--             & 16:41:38.67 & 36:25:37.66 & 1328057012383752704\\
V39$^{1}$    & SR            & 11.92     & 10.40 &--    &--    & 56.0$^{4}$  &--             & 16:41:42.52 & 36:26:55.73 & 1328057145522507648\\
V40$^{1}$    & SR            & 12.10     & 10.67 &>0.01 &>0.01 & 33.0$^{4}$  &--             & 16:41:49.69 & 36:27:48.89 & 1328058657351049344\\
V41          & SR            & 17.18     & 11.98 &>0.13 &>0.09 & 42.5$^{4}$  &--             & 16:41:45.69 & 36:27:57.36 & 1328057935796514560\\
V42$^{1}$    & SR            & 11.88     & 10.39 &>0.02 &>0.03 & 40.0$^{4}$  &--             & 16:41:35.49 & 36:27:27.35 & 1328057871377540224\\
V43          & L             & 12.46     & 11.14 &>0.08 &>0.04 & --          &--             & 16:41:27.07 & 36:28:00.11 & 1328058107595085440\\
V44          & L             & 12.15     & 10.72 &>0.04 &>0.03 & --          &--             & 16:41:39.16 & 36:27:21.86 & 1328057802657985792\\
V45          & L             & 12.64     & 11.41 &>0.02 &>0.03 & --          &--             & 16:41:41.11 & 36:27:22.65 & 1328057905737161344\\
V46          & SX Phe        & 16.04     & 15.41 & 0.15 & 0.14 & 0.052186    & 7574.5437     & 16:41:41.34 & 36:27:05.10 & 1328057905737081856\\  
V47          & SX Phe        & 16.88     & 16.55 & 0.29 & 0.31 & 0.065256    & 7554.5293     & 16:41:38.76 & 36:26:20.90 & 1328057802657690880\\  
V50          & SX Phe        & 16.94     & 16.55 & 0.33 & 0.31 & 0.061754    & 7568.5076     & 16:41:32.45 & 36:28:51.70 & 1328058146249921024\\           
V54$^{2}$    & RRc           & 14.90     & 14.59 & 0.16 & 0.13 & 0.295374    & 7568.4743     & 16:42:07.65 & 36:29:36.53 & 1328058970889601408\\           
V55$^{2}$    & SX Phe        & 17.60     & 17.40 & 0.23 & 0.37 & 0.040505    & 7554.5817     & 16:41:48.47 & 36:26:13.58 & 1328057081095055488\\  
V56$^{2}$    & SX Phe        & 17.21     & 16.88 & 0.18 & 0.28 & 0.024140    & 7572.5045     & 16:41:45.63 & 36:26:54.87 & 1328057149815076736\\  
V57$^{2}$    & W Uma         & 18.60     & 17.98 & 0.52 & 0.46 &0.285416     & 7553.3838$^3$ & 16:41:21.37 & 36:28:17.92 & 1328104669336332544\\
V58$^{2}$    & L             & 13.77     & 12.71 &>0.07 &>0.03 & --          &--             & 16:41:40.70 & 36:27:16.16 & 1328057905737008768\\
V59$^{2}$    & L             & 12.27     & 10.89 &>0.07 &>0.05 & --          &--             & 16:41:33.92 & 36:30:05.51 & 1328058249334814208\\
V60$^{2}$    &$?$            & 14.47     & 13.61 &>0.10 &>0.10 &0.494497     &--             & 16:41:44.53 & 36:27:52.58 & 1328057940089940736\\
\hline
\end{tabular}
\raggedright
\center{\quad \emph{Bl}: RR Lyrae with Blazhko effect.\\
1. Saturated in our IAC images.\\
2. New variable found in the present work. \\
3. Time of minimum. \\
4. Periods determined by \citet{Osborn2017}.}

\end{center}
\end{table*}

\subsection{The search for new variable stars}
\label{search}
\begin{figure}
\includegraphics[width=7cm, height=7cm]{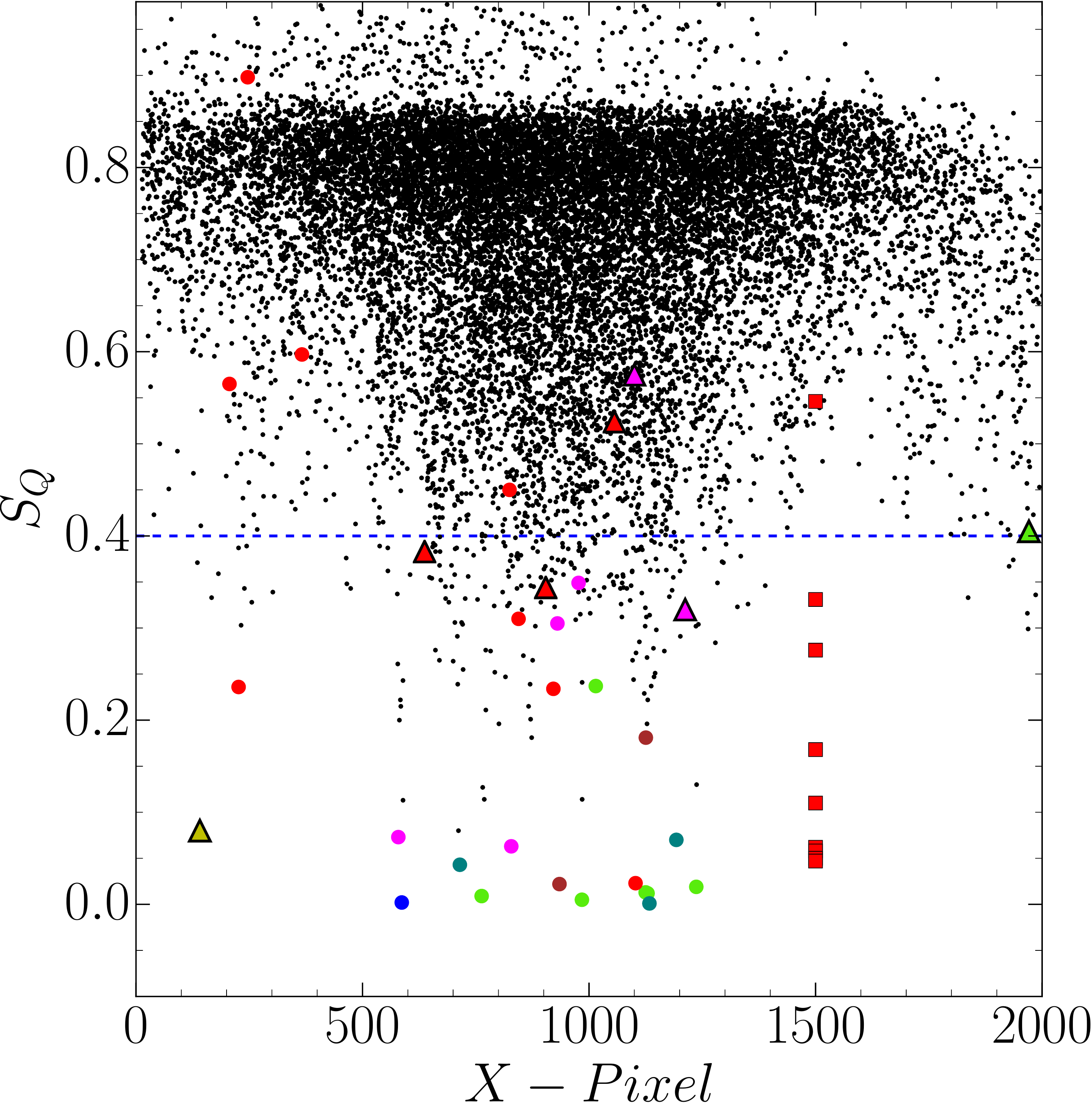}
\caption{Minimum value for the string-length parameter $S_{Q}$ calculated for the 16,280 stars with a light curve in our $V$ reference image, versus CCD X-pixel coordinate. The colour code is as follows: Blue circle corresponds to a RRab star, green circles to RRc stars, brown circles to RRd stars, teal circles to CW stars, red circles correspond to semi-regular variables and magenta circles to SX Phe stars. The triangles correspond to newly discovered variables in this work (one RRc, two SX Phe, three semi-regular and one contact binary). Red squares correspond to semi-regular stars that were measured at Hanle but are saturated in the IAC data and have been assigned an arbitrary pixel coordinate. The dashed blue line is an arbitrary threshold set at 0.4, below which most of the known variables are located. See  $\S$ \ref{search} for a discussion.}
\label{sq} 
\end{figure}

Due to the high quality of our data, we were able to recover 16,280 light curves in $V$ and 16,278 in $I$ of the stars present in our FoV. We carried out a search for new variable stars to enlarge the variable star counts using our data and thereby refine the cluster parameter estimates based on different approaches. 
We used three search methods which will be briefly described below:

\begin{itemize}
\item We split the CMD of M13 into regions where it is common to find variable stars, e.g. at the Instability Strip (IS) in the Horizontal Branch (HB), the Blue Straggler region (BS) and at the tip of the Red Giant Branch (TRGB). We analysed the light curves of the stars in those regions and looked for variability by determining their period (if any) and plotting their apparent magnitudes with respect to their phase. We discovered three new variables with this method: one RRc (V54), and two SX Phe stars (V55 and V56).\\

\item  Another approach used was via the string-length method (\citealt{Burke1970}, \citealt{Dworetsky1983}). Each light curve was phased with periods between 0.02 d and 1.7 d. in steps of $10^{-6}$ d, and the string-length parameter $S_{Q}$  was calculated in each case. The best phasing
is obtained with the true period  and it produces a minimum $S_{Q}$. A plot of the minimum $S_{Q}$ for each star in our collection of light curves (identified by the X-coordinate in the reference image) is shown in Fig. \ref{sq}, where all variables in Table \ref{variables} are identified. We note that most of the variables are located below an arbitrary threshold at 0.4, hence we individually explored each light curve below this value. Using this method, we identified a contact binary (V57) and two long period giant stars (V58 and V59).\\

\item The third method consists in the detection of variations of PSF-like peaks in stacked residual images from which we can see the variable stars blink. All previous known variables were detected, and noted 15 candidates that need confirmation. 
\end{itemize}

\subsection{The RR Lyrae stars}

\begin{figure*}
   \centerline{\includegraphics[width=18cm, height=16.0cm]{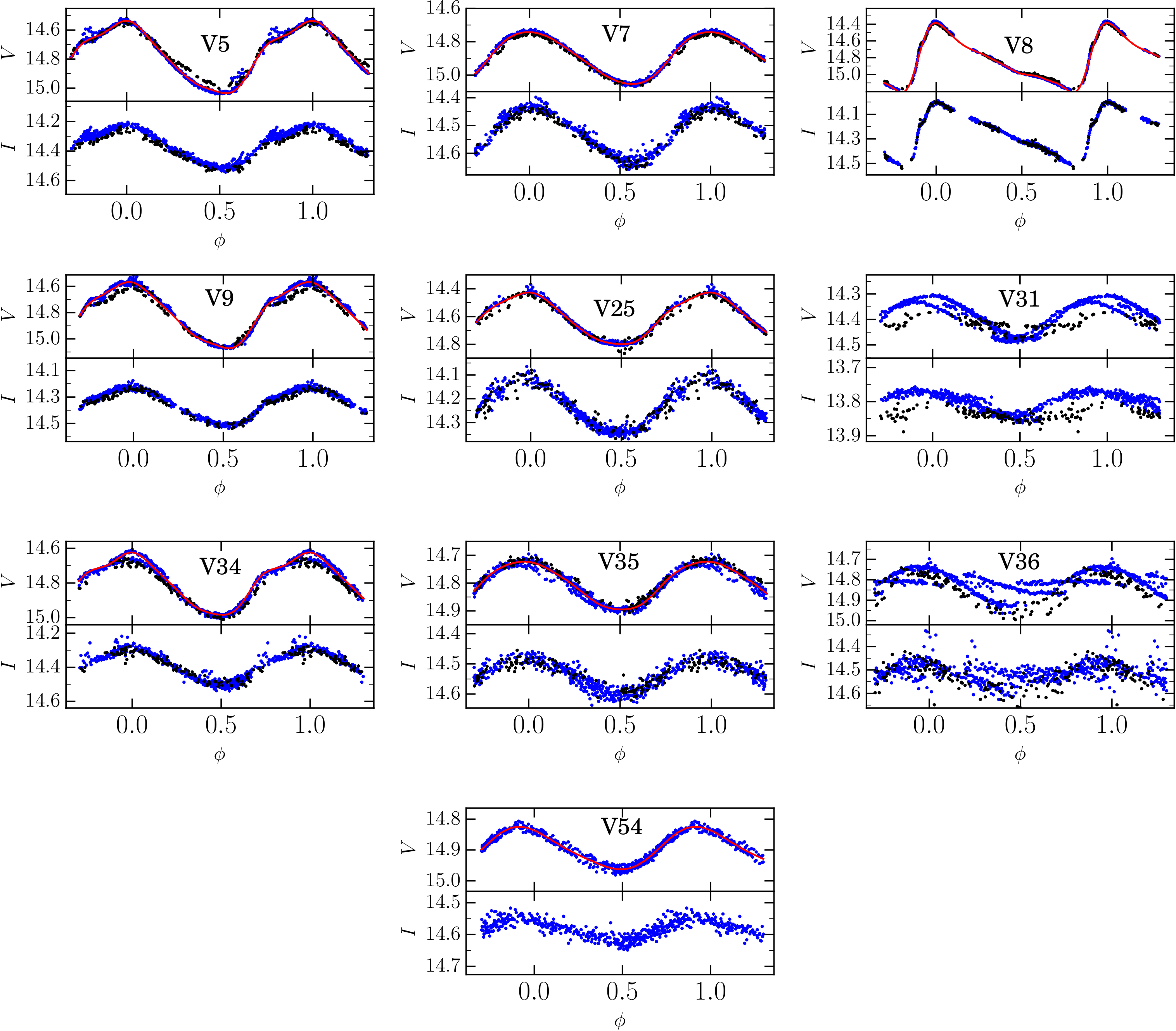}}
 \caption{Light curves in $V$ and $I$ filters of RR Lyrae stars in M13. The light curves are a combination of the data obtained at IAC (blue symbols) and at Hanle (black symbols). The continuous red line represents the Fourier fit. Note that the vertical scale is not the same for all plots. Note the double-mode nature of V31 and V36.}
\label{LC_RR}
\end{figure*}

\subsubsection{RRab and RRc stars}
\label{RRabc}

With the newly found  RRc star (V54), the RR Lyrae population of M13 consists of one RRab (V8), 7 RRc, and two RRd (V31 and V36). Their $VI$ light curves are shown in Fig. \ref{LC_RR}.

\subsubsection{Multi-frequency RR Lyrae stars}
\label{RRd}

The star V36 is a well known multi-frequency variable.
\cite{Kopacki2003} identified three frequencies, very close to each other and concluded that the star belongs to a group of RR Lyrae with non-radial modes and period ratios larger than 0.95 \citep{Olech1999}.
We have attempted to reproduce the observed $V$ and $I$ light curves of V36
as double-mode oscillations of two close frequencies,
with the adopted frequencies determined by minimization of the
squared residuals between the model and the observations. We found the two frequencies and amplitudes reported in Table \ref{variables}.

\begin{figure*}
   \centerline{\includegraphics[width=16cm, height=6.0cm]{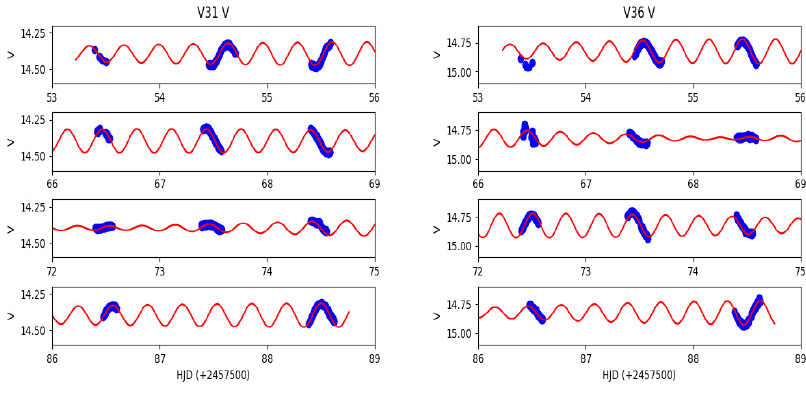}}
 \caption{V31 and V36 data fitted with a two frequency model.}
\label{V36fit}
\end{figure*}

These periods correspond very closely to the first two frequencies found by \cite{Kopacki2003}. Fig. \ref{V36fit} shows the $V$-band model with $P_1 =0.31596$ d and $P_2 =0.30424$ d. With the precision of our photometry we were unable to identify a third period $P_3$.
Similar periods and amplitudes were found from the analysis of the $I$-band data.

The star V31 is also a double-mode star that apparently went unnoticed by \cite{Kopacki2003}. The $VI$ light curves of V31 in Fig. \ref{LC_RR} suggest the presence of more than one period. Similar to V36, we identified in V31 two periods; $P_1 = 0.32904$ d and $P_2 = 0.31936$ d. Both V31 and V36 have a large period ratio $P_2/P_1 \sim 0.97$, implying that at least one of the modes is non-radial. The two-frequency model fitting for V31 is shown in Fig. \ref{V36fit}.

\subsubsection{Bailey diagram and Oosterhoff type}
\label{baileyDiagram}

\begin{figure}
\includegraphics[width=6cm, height=10cm]{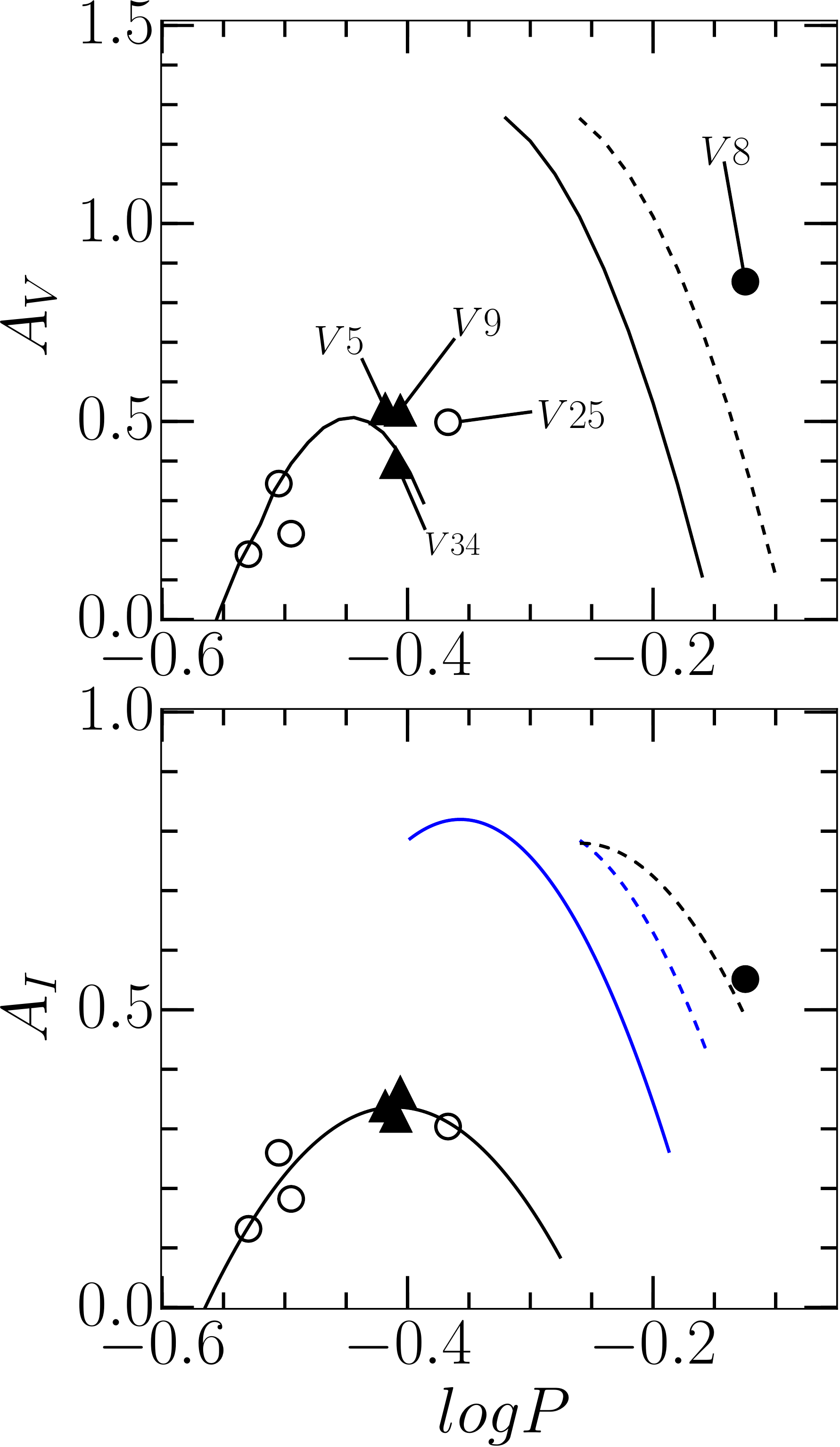}
\caption{Bailey diagram for M13. Filled and open circles represent RRab and RRc stars, respectively. Triangles correspond to stars with Blazhko modulations. For a more detailed discussion, see $\S$ \ref{baileyDiagram}.}
\label{figBailey} 
\end{figure}

The period-amplitude plane for RR Lyrae stars, also known as the Bailey diagram, is shown in Fig. \ref{figBailey} for the \emph{VI} band passes. The periods and amplitudes are listed in Table \ref{variables}. In most cases, we took the amplitudes corresponding to the best fit provided by the Fourier decomposition of the light curves. In cases where the light curve showed Blazhko effect (V5, V9 and V34), the maximum amplitude was measured and the star was plotted with a triangular marker. The continuous and dashed black lines in the top panel of  Fig. \ref{figBailey} are
the loci for unevolved and evolved stars according to \citet{Cacciari2005}.
The black parabola, was obtained by \citet{Kunder2013b} 
for RRc stars in 14 OoII clusters. In the bottom panel, the black dashed
locus was found by \citet{Arellano2011} and \citet{Arellano2013} for the OoII clusters NGC 5024 and NGC 6333 respectively. The
black parabola was obtained in the present work using a least-squares fit
with the RRc stars. The blue solid and segmented loci for unevolved and
evolved stars respectively are from \citet{Kunder2013a}.
The positions of the RRab and RRc stars in this diagram are consistent with the definition of a OoII type globular cluster.\\

\subsection{The W Virginis or CW stars}

\begin{figure*}
   \centerline{\includegraphics[width=16cm, height=8.0cm]{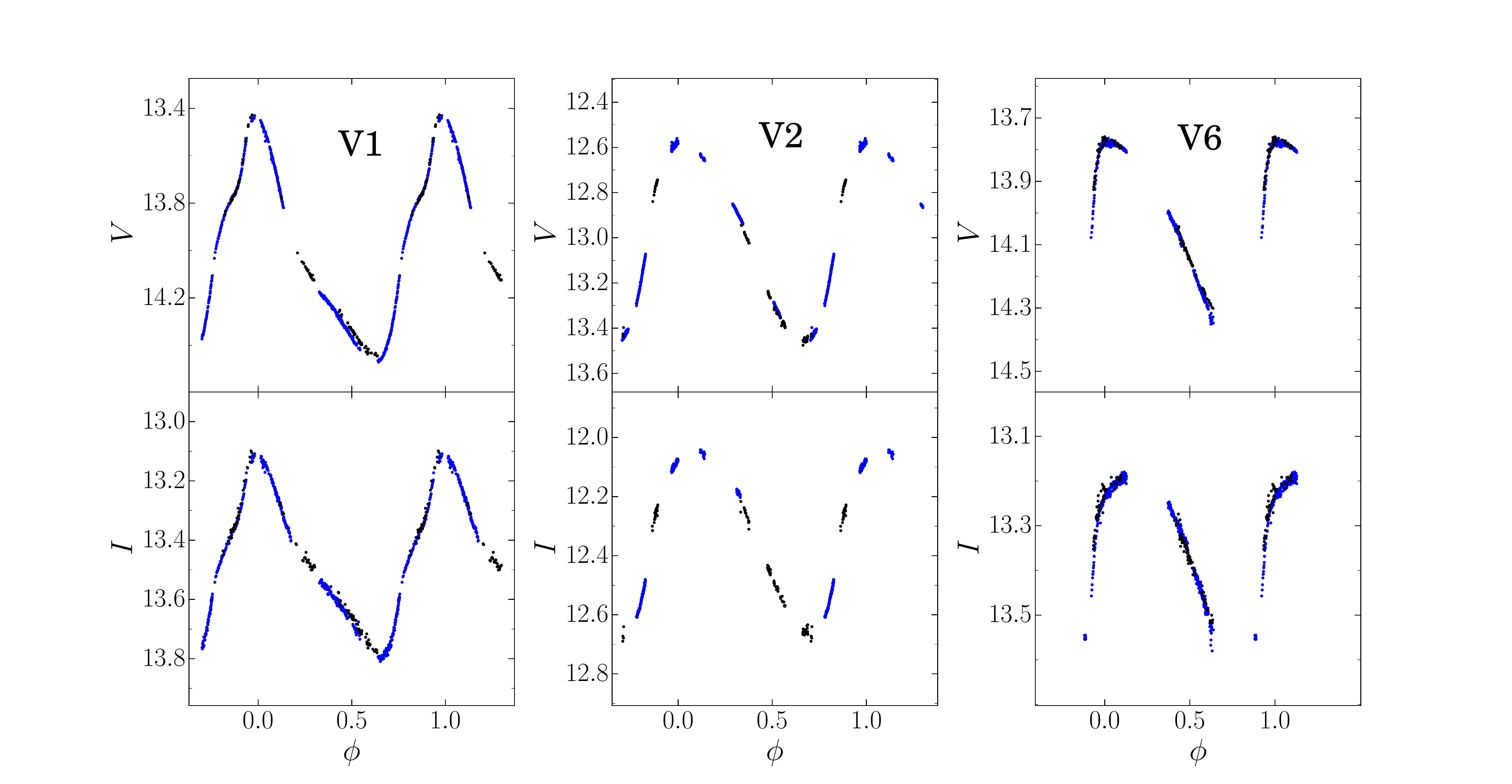}}
 \caption{Light curves in $VI$ filters of the three CW stars known in M13. The light curves are a combination of the data obtained at IAC (blue dots) and at Hanle (black dots). Note that the scale is not the same for all plots.}
\label{LC_CW}
\end{figure*}

Three W Virginis stars are known in M13; V1, V2 and V6. Their light curves are shown in Fig. \ref{LC_CW} phased with the periods reported in Table \ref{variables}.

\subsection{The SX Phe stars}

\begin{figure*}
   \centerline{\includegraphics[width=16cm, height=10.0cm]{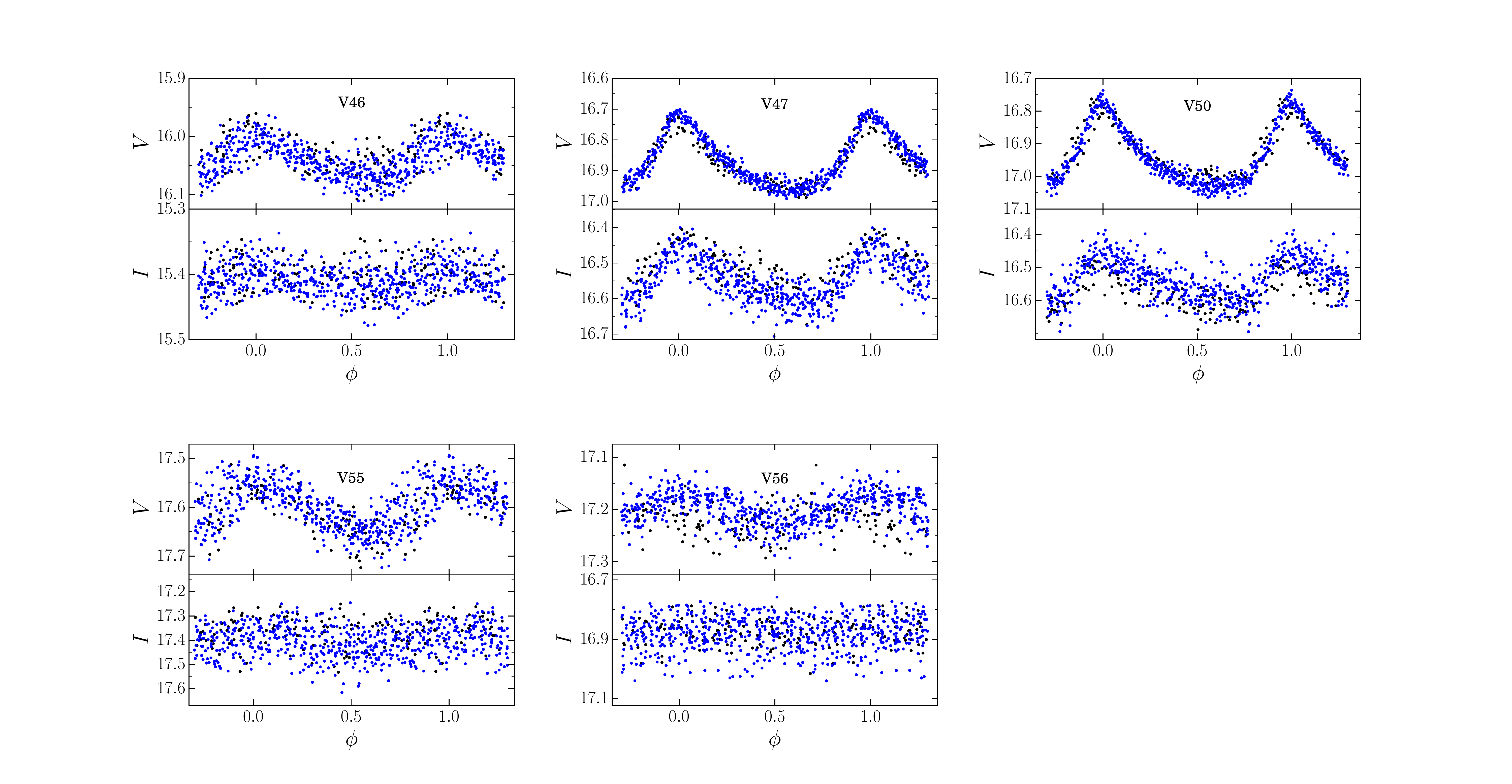}}
 \caption{Light curves in $V$ filter of the SX Phe in M13. The light curves are a combination of the data obtained at IAC (blue dots) and at Hanle (black dots). Note that the vertical scale is not the same for all plots.}
\label{LC_SX}
\end{figure*}

Four SX Phe stars are known in M13; V37, V46, V47 and V50. Their frequency spectra have been studied in detail by \cite{Kopacki2005}. 
Further comments are needed for these stars. V37 is located very near two brighter stars and they are not properly separated in our images (see the identification chart in Fig. \ref{M13_CHART}b near the top). As a consequence we could not retrieve the light curve of the star. We have adopted the $V$ and $V-I$ values of \cite{Kopacki2005} to include the star into the discussions below. V46 was reported by \cite{Kopacki2005} as a $V = 17.2$ magnitude star with at least two excited frequencies, and probably three. The coordinates given by \cite{Kopacki2005} point to the star marked in Fig. \ref{M13_CHART} and we found the same period calculated by these authors. The light curve is shown in Fig. \ref{LC_SX}. We note that the mean $V$ magnitude is 16.04, i.e. more than one magnitude brighter than the one reported by \cite{Kopacki2005}. Also, we were unable to find any secondary periodicity in V46, hence we do not confirm its multi-frequency nature. Likewise, for V47 we find a mean $V=16.88$, versus 17.12 of \cite{Kopacki2005}.
For V50, the mean $V$ magnitudes in both studies match within 0.01 mag.
We stress, however, that the identifications and periods found by us, do coincide with those of \cite{Kopacki2005} within a few millionths of a day, which rules out possible misidentifications.

In the present paper we have identified two more variables that according to their period, light curve shape and position on the CMD are classified as SX Phe stars; V55 and V56. The light curves of the SX Phe stars are shown in Fig. \ref{LC_SX}. Their position in the CMD and their use as distance indicators will be discussed in $\S$ \ref{sxphe}.

\subsection{The new contact binary V57}
\label{V57}

\begin{figure}
   \centerline{\includegraphics[width=7cm, height=12.0cm]{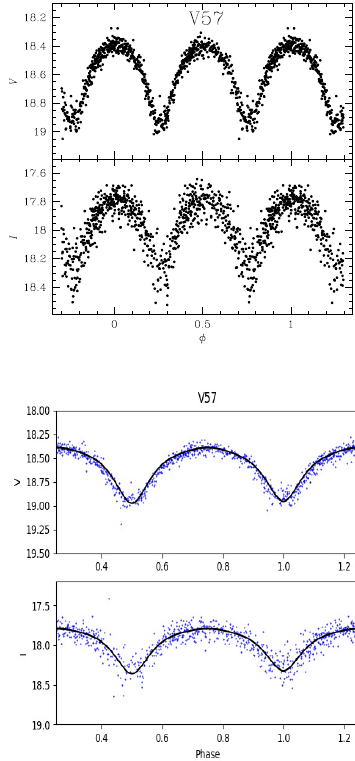}}
 \caption{Light curves of the contact binary V57. The bottom panels show the adopted model fitted to the observations. See $\S$ \ref{V57} for a detailed discussion.}
\label{algolV57}
\end{figure}

We show in Fig. \ref{algolV57} the $VI$ light curves of the new binary discovered in our data. From the time of the minima in our light curves we derived the ephemeris for the primary minimum: HJD= $2457553.38380 ~+~ 0^d.285416 ~E. $

The morphology of the light curves of this short-period binary is similar
to the light curves of contact systems. In order to derive some information on the physical parameters of the system, we have modeled the $VI$ light curves with the code BinaRoche, some details of which are described in \citet{Lazaro2009}, with the two light curves used simultaneously in the fit. For this work, the code uses surface fluxes from the library of theoretical stellar spectra of \citet{Lejeune1998} for [Fe/H]= -1.50. As we don't have radial velocity curves, the masses of the stellar components, which are essential parameters in the model, must be estimated by some indirect reasoning. All that we have is the colour $(V - I)$ of the binary, which is a rather scarce information, but even then we can reproduce the observed light curves with a reasonable model of the system.\\
The observed $V$ magnitude and colour in the maxima of the light curves are:
$$ V_{bin.,max.}= 18.40  ~~ , ~~ (V - I)_{bin.,max.}= 0.60 $$

 From the observed $(V-I)$ colour out of eclipse, and assumed $E(B-V)= 0.02$,
 we derive the intrinsic colour of the system out of eclipse  ~$(V - I)_{bin.,o}= 0.57$. \\
 
From this combined colour of the binary, we attempted to derive the intrinsic colours $(V - I)_{1,2}$ of the binary components. This can be done if the relative contribution of the stars to the total light in $V$ and $I$ are known, which can be derived from the model fit, and a $(V - I) - T_{\rm eff}$ calibration. For that purpose we have adopted the relation of \citet{Casagrande2010},
 valid for the ranges in metallicity and colour suitable to our data.\\
In order to calculate a model of the system, it is necessary to give values to the masses of the components. This has been performed including the relation between $M/M_\odot$ and  log$~T_{\rm eff}$ from the Padova's isochrone adopted for the cluster, and the relation $(V - I) - T_{\rm eff}$ in the simultaneous fit of the $V$, $I$ light curves. \\
After some trials, the filling factors of the stars have been fixed to a contact solution, while the parameters $M_1, q= M_2/M_1 , T_{\rm eff,1},  T_{\rm eff,2}$, $i$ (inclination angle), $A_{b,1}, A_{b,2}$ (bolometric albedo coefficients), $\beta_1, \beta_2$ (gravity-darkening coefficients), are declared free and optimized.\\
From the best fit, the main derived parameters are:
$$ M_1= 0.87 ~\pm~ 0.02 ~M_\odot  ~~ , ~~ q= M_2/M_1= 0.72 ~\pm~ 0.02 $$
$$ T_{\rm eff,1}= 6400 ~\pm~ 100 ~K ~~ , ~~ T_{\rm eff,2}= 6285 ~\pm~ 100 ~K $$
$$ R_{equiv.,1} (R_\odot)\simeq  0.85  ~~ , ~~R_{equiv,2} (R_\odot)\simeq  0.73 $$
$$ i = 79^o ~\pm~ 2^o $$
$$A_{b,1}= 0.5, A_{b,2}= 0.5, \beta_1= 0.08, \beta_2= 0.08 $$

This solution suggests that V57 can be a new W UMa type binary.\\ 
 
The adopted model gives a distance to the system ~ $d$= 6.95 ~kpc. (for $E(B-V)$ = 0.02 mag). This value is in good agreement with those derived by the other methods, and gives some confidence on the adopted model for the system.\\
From the observed $V$ magnitude and the distance of the model, the absolute $V$ magnitude of the system is ~$M_V= +4.13$, in good agreement with that expected from the P-L-colour relation of \citet{Rucinski1995} for W UMa systems in globular clusters:
$$ M_V = -4.43 ~log P + 3.63 (V - I_c) - 0.31 ~ (\sigma= 0.29 ) = +4.17. $$

\begin{figure}
   \centerline{\includegraphics[width=10cm, height=9.0cm]{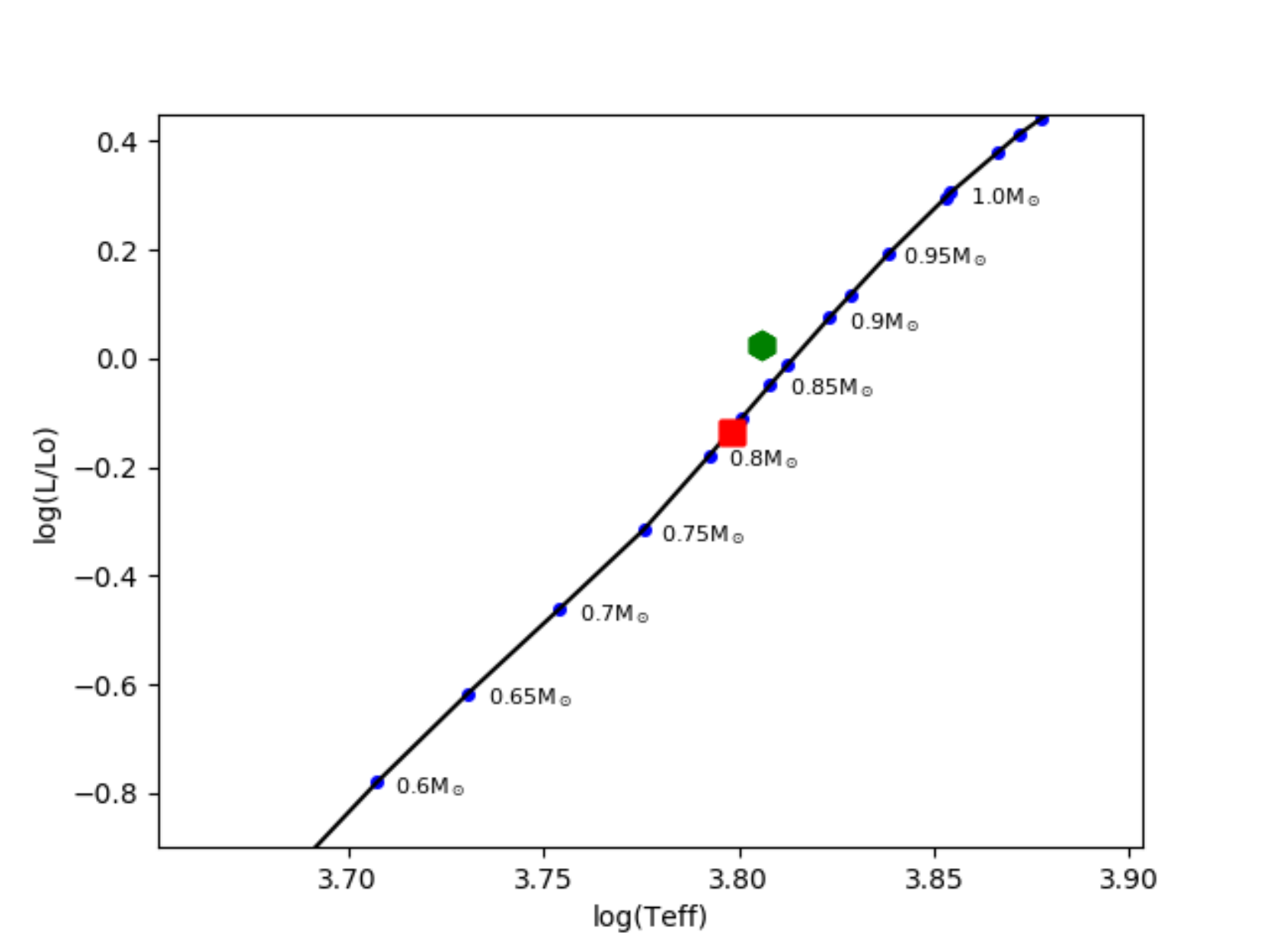}}
    \caption{Position of the two components of V57 in the $\rm log~T_{\rm eff} - log~L$ plane along with theoretical isochrone values. The green hexagon and red square correspond to the primary and secondary components respectively. See $\S$ \ref{V57} for a discussion.}
\label{V57planes}
\end{figure}

 In Fig. \ref{algolV57} the observed and model light curves are shown. In Fig. \ref{V57planes} we show  the location of both stellar components in the $log T_{\rm eff} - log \rm L$ diagram, together with the theoretical isochrone values. In this diagram, the green hexagon and red square correspond to the primary and secondary component respectively. The derived values of $T_{\rm eff}$ (mean surface temperature) and $R_{equiv}$ (equivalent volume radius) have been used to calculate the luminosity of the stellar components shown in the diagram.\\
 
As the model relies on very limited data, we present this solution as a
preliminary model of the new binary. More multi-colour photometric
light curves, and radial velocity curves, would be necessary to collect and analyse before we can be confident on the nature of the system. \\

\subsection{The giant variable SR or L stars}
\label{SR_CAND}

\begin{figure}
   \centerline{\includegraphics[width=7cm, height=8.0cm]{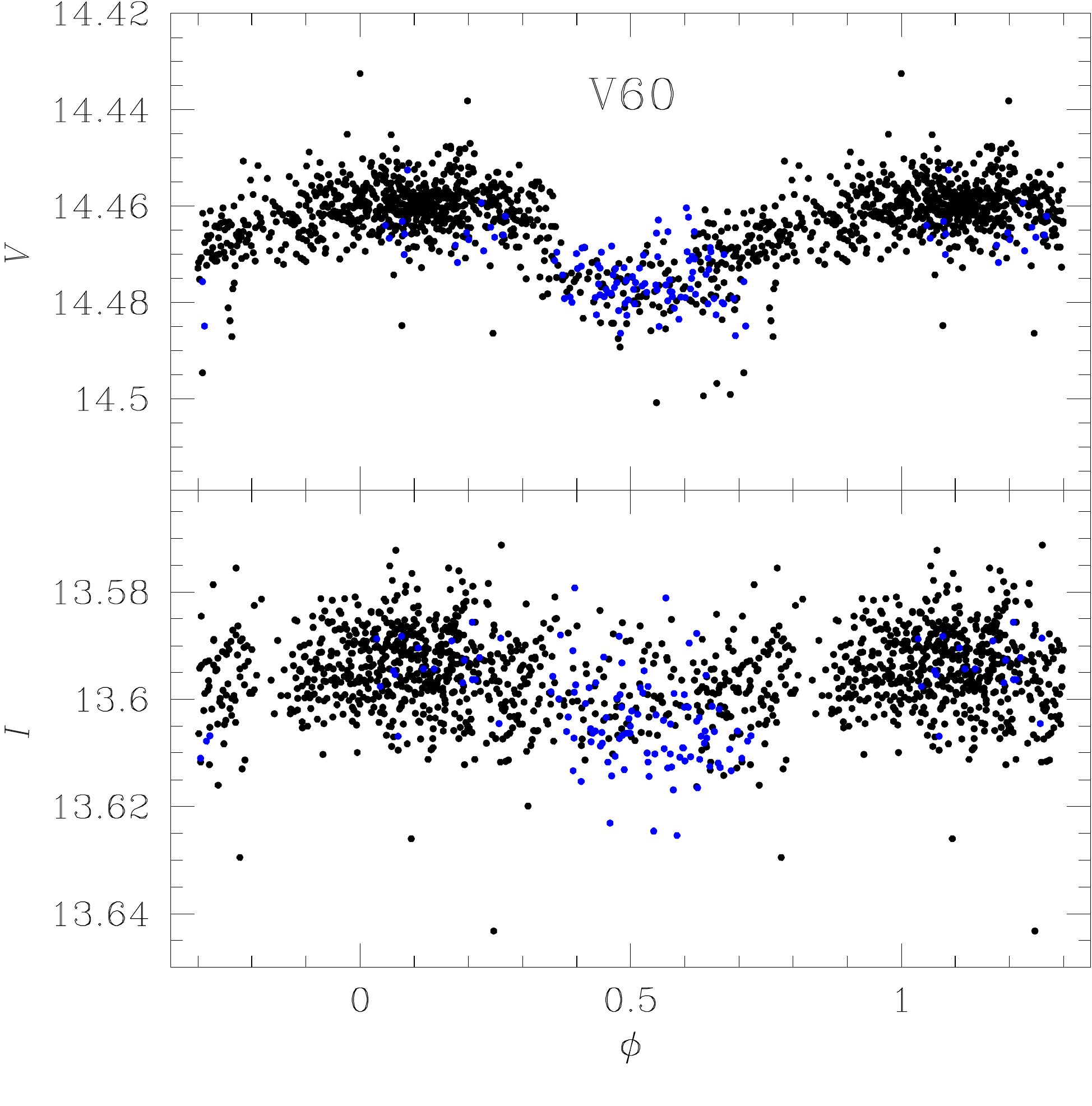}}
 \caption{Short-term (0.494497 d) variations  of the peculiar semi regular variable V60. Blue symbols are data from 2014 and have been shifted in phase and magnitude  to match black symbols from 2016, in order to compensate for the obvious long-term variations seen in Fig. \ref{LC_SR}. See $\S$ \ref{SR_CAND} for a detailed discussion.}
\label{V60}
\end{figure}

\begin{figure*}
   \centerline{\includegraphics[width=18cm, height=10.0cm]{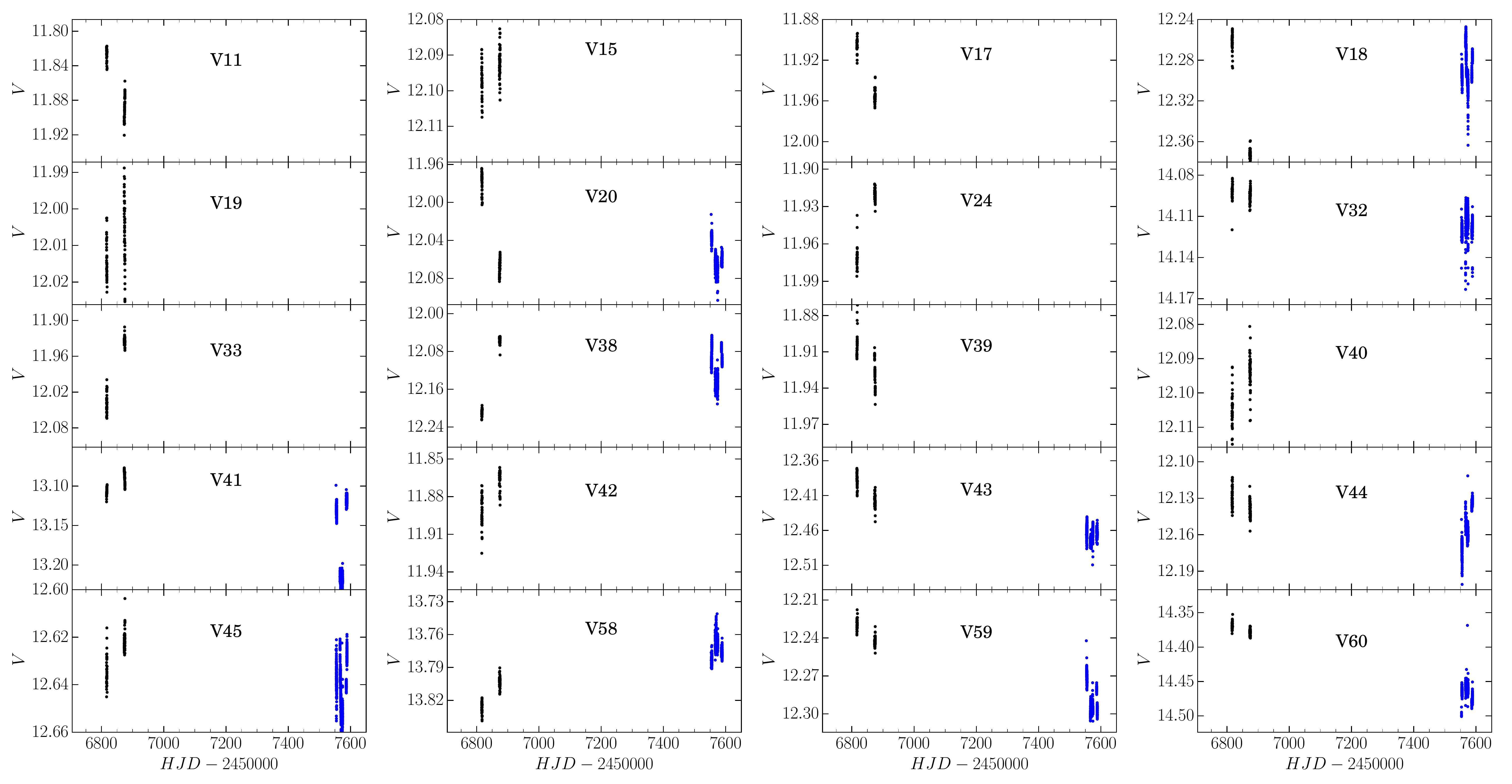}}
 \caption{Light curves in $V$ filter of the SR and L type variables in M13. The light curves are a combination of the data obtained at IAC (blue dots) and at Hanle (black dots). 
 Note that the scale of the $V$-axis is not the same for all plots.}
 \label{LC_SR}
\end{figure*}

Numerous variable bright giants are known in this cluster.
Unfortunately a few of them are saturated in our IAC images.
A plot of the $V$ magnitude vs. HJD reveals the long term variations of two previously unreported variables which we have called V58 and V59 and  classified as L-type. Also, a new variable V60 is found between the HB and the RGB, not far from V32, which was considered by \citet{Osborn2017} to be a non-typical SR (see Fig. \ref{V60}). V60 displays a long-term variation (see Fig. \ref{LC_SR}) but a short-term variation with a period of 0.494497 d is also evident in the data of 2014 and 2016, after the 2014 data are adequately shifted to the 2016 data. The short term variations may be due to the binary nature of the stars.

\begin{figure}
   \centerline{\includegraphics[width=8.0cm, height=8.0cm]{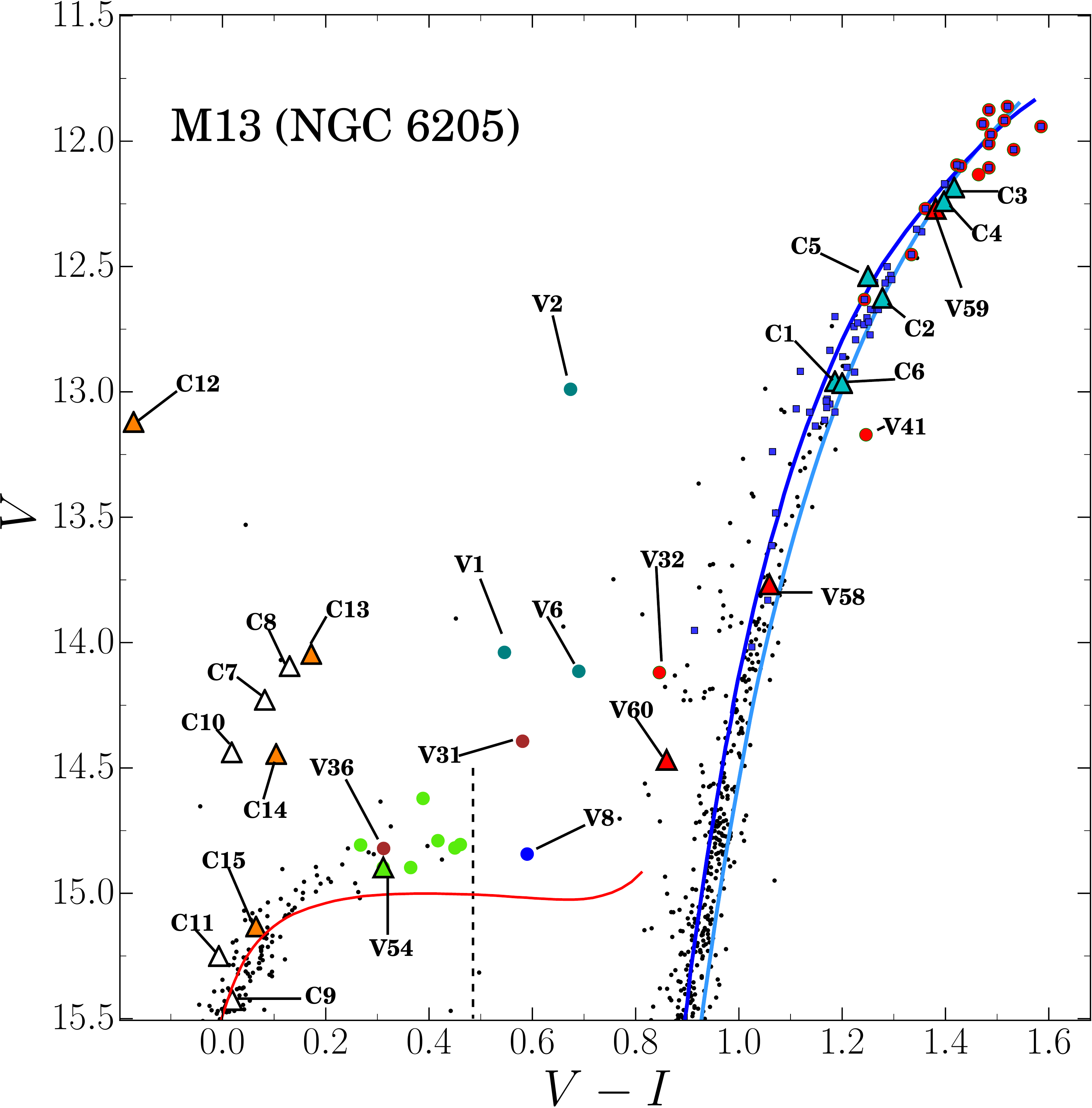}}
 \caption{Upper section of CMD of M13 displaying, along with some well established variables, the 15 variable stars candidates identified in the FoV of the cluster by blinking techniques. These candidates are represented by the empty, orange and cyan triangles and labeled as C stars, and their light curves are shown in Fig. \ref{LC_cand}. The empty triangles correspond to five variable stars candidates that might not be truly cluster members and the four orange triangles correspond to non-members. The full CMD of M13 is in Fig. \ref{CMD_6205}. See $\S$ \ref{SR_CAND} for a discussion.}
\label{osbornM13}
\end{figure}

\begin{figure*}
   \centerline{\includegraphics[width=18cm, height=8.0cm]{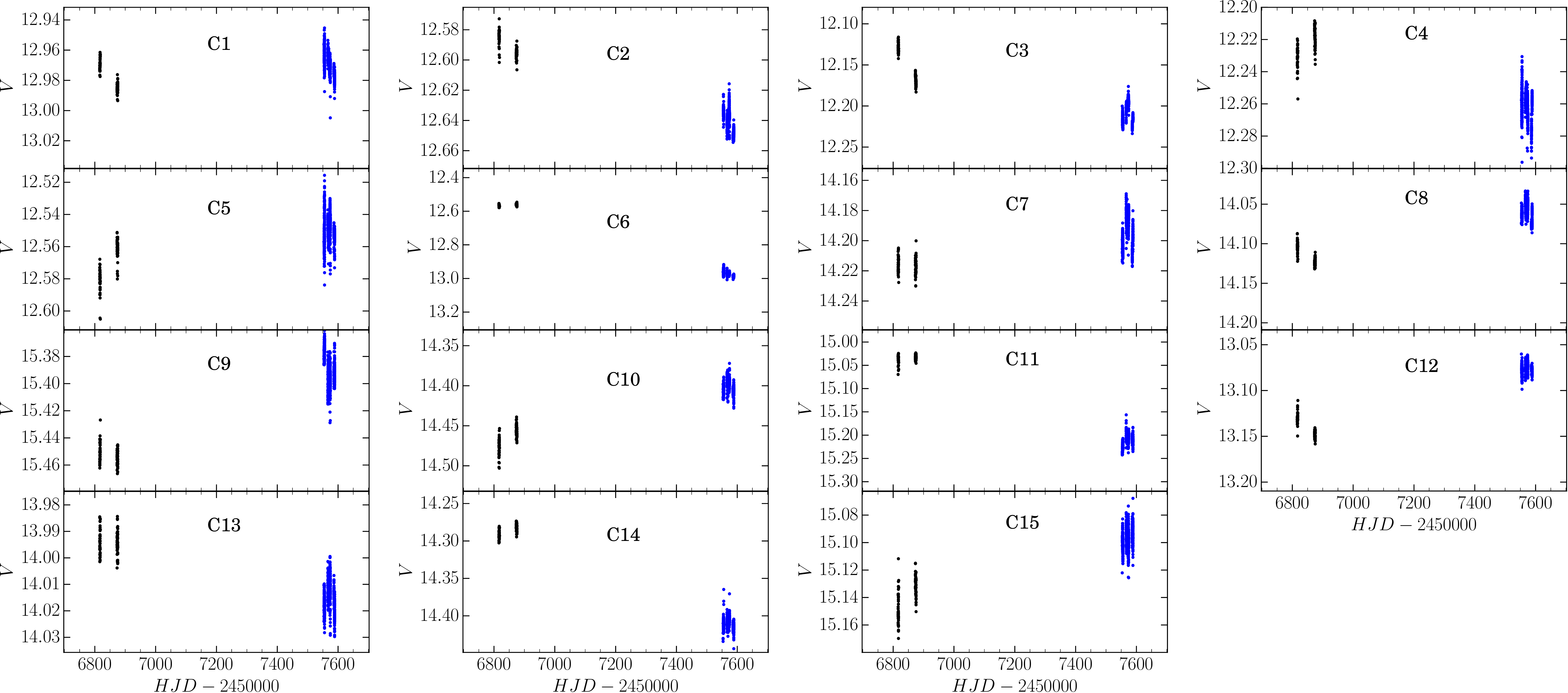}}
 \caption{Light curves of 15 stars in the field of M13 not previously  known as variables. Colours are as in Fig. \ref{LC_SR}. We have retained them as candidates although their variations are rather conspicuous. A blow up of the blue symbols and the $I$ light curves confirm the shorter term variations (not illustrated).}
 \label{LC_cand}
\end{figure*}

A blinking inspection of a series of images, suggested light variations of a group of stars not previously detected as variables. The plots of the $V$ magnitude as a function of HJD show a conspicuous variation in most of them (Fig. \ref{LC_cand}). We confirm that their mid-term variations are comparable to those observed in well established SR variables (Fig. \ref{LC_SR}). We have refrained from assigning them a variable star number, which shall be assigned in the Catalog of Variable Stars in Globular Clusters (CVSGC) once the variation is confirmed. We will denote these stars with a C designation (for candidates). Fig. \ref{osbornM13} displays the CMD of the cluster above the HB to illustrate the position of these variable candidates. Some of them are clearly in the RGB and are cluster members, judging from their proper motions and/or radial velocities according to  \textit{Gaia}-DR2, which shall be further discussed in $\S$ \ref{gaia_members}. Five of these candidates are found to be  above and/or to the blue side of the HB; C7, C8, C9, C10 and C11 (empty triangles in Fig. \ref{osbornM13})
and their proper motions and quality parameters reported in \textit{Gaia} make their membership dubious. The four candidates C12, C13, C14 and C15 are definitely not cluster members.

\begin{table*}
\begin{center}
\caption{Data of variable stars candidates in M13 in the FoV of our images. See discusion in $\S$ \ref{SR_CAND}.}
\label{candidatas}
\begin{tabular}{lcccccccc}
\hline
Star ID & Type & $\overline{V}$ (mag) & $\overline{I}$ (mag)   & $A_V$ (mag)  & $A_I$  (mag)   & $\alpha$ (J2000.0)  & $\delta$  (J2000.0) & Membership        \\
\hline

C1  &SR &12.97 &11.78  &>0.04 &>0.04  & 16:41:32.33 & +36:27:34.52 & yes \\
C2  &SR &12.63 &11.36  &>0.07 &>0.07  & 16:41:34.33 & +36:30:13.03 & yes \\
C3  &SR &12.20 &10.79  &>0.10 &>0.09  & 16:41:34.75 & +36:27:59.27 & yes \\
C4  & SR  &12.25 &10.87  &>0.07 &>0.06  & 16:41:34.80 & +36:27:19.38 & yes  \\
C5  & SR  &12.55 &11.31  &>0.07 &>0.05  & 16:41:35.68 & +36:26:48.61 & yes  \\
C6  & SR  &12.87 &11.69  &>0.44 &>0.50  & 16:41:39.73 & +36:26:37.79 & yes  \\
C7  & irregular  &14.20 &14.09  &>0.06 &>0.08  & 16:41:39.77 & +36:28:06.43 & uncertain \\
C8  & irregular  &14.07 &13.90  &>0.09 &>0.13  & 16:41:40.41 & +36:28:09.81 & uncertain \\
C9  & irregular  &15.40 &15.36  &>0.09 &>0.13  & 16:41:42.84 & +36:31:00.93 & uncertain \\
C10 & irregular  &14.41 &14.36  &>0.11 &>0.07  & 16:41:43.56 & +36:27:11.02 & uncertain \\
C11 & irregular  &15.18 &15.19  &>0.20 &>0.15  & 16:41:44.38 & +36:26:21.50 & uncertain \\
C12 & irregular  &13.08 &13.23  &>0.09 &>0.14  & 16:41:33.65 & +36:26:07.54 & no  \\
C13 & irregular  &14.01 &13.81  &>0.04 &>0.07  & 16:41:34.75 & +36:29:13.74 & no  \\
C14 & irregular  &14.39 &14.29  &>0.14 &>0.08  & 16:41:39.03 & +36:27:09.33 & no  \\
C15 & irregular  &15.10 &15.01  &>0.08 &>0.10  & 16:41:45.60 & +36:26:37.42 & no  \\
\hline
\end{tabular}
\raggedright
 \end{center}
\end{table*}

In their paper, \citet{Osborn2017} performed a thorough analysis of the most luminous variables in the RGB of M13. They estimated their cycle times between 30 and 90 days and point out that some of them may have multiple periods. These authors also find that the magnitude range of the variation and period increase with luminosity. This result forecasts the difficulty of finding variable stars in the lower parts of the RGB. 

We call attention to these variables on the RGB, which are confirmed members from their \textit{Gaia} proper motions and/or radial velocities. This also confirms that virtually all stars in M13 above $V < 12.5$ mag and with ($V-I$) greater than about 1.2 mag do present mid- to long-term variations.

Given the limited time distribution of our data, we have not attempted to measure periods or characteristic cycle times for the stars in the RGB. For the previously known SR stars, the given periods in Table \ref{variables} are those of \citet{Osborn2017}.

\subsection{Comments on the RR Lyrae stars  identified in \textit{Gaia}-DR2 in the field of M13}

\citet{Clementini2018} reports the detection and characterisation of 140,784 RR Lyrae stars by a Specific Object Study pipeline in \textit{Gaia}-DR2 all over the sky. The sample of confirmed RR Lyrae stars includes those found in 87 globular clusters including the cluster being studied in the present work. They provide the multi-band time series data (in the \textit{Gaia} photometric system) and derive from them the characteristic physical parameters of the listed RR Lyrae stars. From the stars listed by \citet{Clementini2018}, we selected those inside the tidal radius of M13, 23 arcmin according to the Catalogue of Milky Way Stellar Clusters by \citet{Kharchenko2013}. We found the eighteen stars listed in Table \ref{clementini}. This table includes the results of our analysis of their variability from our data. It can be seen that only the four previously known RR Lyrae stars were confirmed with our data, no variability was found in seven stars, six stars were too faint for our observations or we were unable to build a reliable light curve, and one is out of the FoV of our images.

\begin{table*}
\caption{Results of our analysis of the RR Lyrae stars reported by \citep{Clementini2018} in M13. The types and periods were derived by \citep{Clementini2018}, the remaining columns were extracted from \textit{Gaia}-DR2.}
\label{clementini}
\begin{tabular}{lllllll}
\hline
\textit{Gaia}-DR2 source & Type & Gmag &  P (days) & RA & DEC & Result of our analysis \\
\hline
1328057179882276480 & RRc  & 14.7406 &  0.381796 & 16:41:46.37 & +36:27:40.0 & Previously known as V5, P=0.381784d.         \\
1328057940097496064 & RRab & 16.7973 & 0.414622 & 16:41:42.49 & +36:28:26.9 & No variability detected in our observations. \\
1328057184175404544 & RRc  & 14.7416 &  0.392737 & 16:41:46.27 & +36:27:37.9 & Previously known as V9, P=0.392724d.         \\
1328057905737005696 & RRab & 16.4233 &  0.691576 & 16:41:39.91 & +36:27:48.3 & No variability detected in our observations. \\
1328057321618952064 & RRab & 19.9812 &  0.669728 & 16:41:25.43 & +36:24:47.1 & Fainter than the limit of our observations.  \\
1328057940089939840 & RRc  & 16.7692 &  0.342303 & 16:41:44.43 & +36:27:54.4 & Unable to obtain a reliable light curve.     \\
1328057768297931904 & RRc  & 14.8746 &  0.312663 & 16:41:37.12 & +36:26:28.9 & Previously known as V7, P=0.312668d.         \\
1328054297963963392 & RRab & 18.5601 &  0.644475 & 16:42:13.42 & +36:23:21.5 & Out of the field of our observations.        \\
1328056634424210560 & RRab & 19.8033 &  0.615953 & 16:41:33.90 & +36:24:47.1 & Fainter than the limit of our observations.  \\ 
1328055397466801664 & RRab & 19.9956 &  0.614019 & 16:42:03.18 & +36:26:22.1 & Fainter than the limit of our observations.  \\
1328057424694539776 & RRab & 18.9557 &  0.471707 & 16:41:28.79 & +36:25:42.4 & No variability detected in our observations. \\
1328056703145793024 & RRab & 19.1703 &  0.570260 & 16:41:47.80 & +36:24:15.0 & No variability detected in our observations. \\
1328055397466713728 & RRab & 20.3268 &  0.583670 & 16:42:04.43 & +36:26:10.3 & Fainter than the limit of our observations.  \\
1328056462625419776 & RRab & 20.3024 &  0.516029 & 16:41:29.53 & +36:22:49.5 & Fainter than the limit of our observations.  \\
1328057733938742528 & RRab & 18.7482 &  0.591195 & 16:41:25.72 & +36:28:08.6 & No variability detected in our observations.   \\
1328058077530125568 & RRab & 14.7583 &  0.750291 & 16:41:32.65 & +36:28:02.1 & Previously known as V8, P=0.750303.          \\
1328056939364066176 & RRab & 18.8948 &  0.481375 & 16:41:41.01 & +36:24:52.6 & No variability detected in our observations. \\
1328057012383745152 & RRab & 17.8061 &  0.703458 & 16:41:39.55 & +36:25:45.1 & No variability detected in our observations.     \\
\hline
\end{tabular}
\end{table*}

\section{RR Lyrae stars: [Fe/H] and Mv from light curve Fourier decomposition}
\label{decomposition}
Since the light curves of RR Lyrae stars are periodic, it is natural to use Fourier series to describe them with the following equation:

\begin{equation}
    m(t) = A_{0} + \sum_{k=1}^{N}A_{k}cos(\frac{2\pi}{P}k(t-E_{0}) + \phi_{k})
	\label{eq:quadratic}
\end{equation}

\noindent
where $m(t)$ is the magnitude at time $t$, $P$ is the period of pulsation, and $E_0$ is the epoch. When calculating the Fourier parameters, we made use of a least-squares approach to estimate the best fit for the amplitudes $A_k$ and phases $\phi_k$ of the light curve components.  The phases and amplitudes of the harmonics in Eq. \ref{eq:quadratic}, i.e. the Fourier parameters, are defined as $\phi_{ij} = j\phi_{i} - i\phi_{j}$, and $R_{ij} = A_{i}/A_{j}$. For the RR Lyrae stars, their Fourier coefficients are listed in Table \ref{tab:fourier_coeffs}.

Over the last few years, our group has consistently used well tested calibrations and zero points to study over 30 globular clusters in the Galaxy  \citep{Are2017}. 
Below, we list the specific set of equations used in this work. 
To calculate the metallicity and absolute magnitude of the RRab stars, we employed the calibrations of \cite{Jurcsik1996} and \cite{Kovacs2001}, respectively:\\

\begin{equation} 
{\rm [Fe/H]}_{J} = -5.038 ~-~ 5.394~P ~+~ 1.345~\phi^{(s)}_{31},
\label{eq:JK96}
\end{equation} 

\begin{equation} 
M_V = ~-1.876~\log~P ~-1.158~A_1 ~+0.821~A_3 + K.
\label{eq:ḰW01}
\end{equation} 

\noindent  
Note that the equation for the metallicity is given in the Jurcsik-Kov\'acs scale, but using [Fe/H]$_{J}$ = 1.431[Fe/H]$_{ZW}$ + 0.88 \citep{Jurcsik1995}, it can be transformed to the standard Zinn-West scale \citep{Zinn1984}. We have adopted the value for $K$ = 0.41 from \citet{Arellano2010}. \\

In the case of the RRc stars, we employed the calibrations given by  \cite{Morgan2007} and \cite{Kovacs1998}, respectively:


$${\rm [Fe/H]}_{ZW} = 52.466~P^2 ~-~ 30.075~P ~+~ 0.131~\phi^{2(c)}_{31}$$
\begin{equation}
~~~~~~~	~-~ 0.982 ~ \phi^{(c)}_{31} ~-~ 4.198~\phi^{(c)}_{31}~P ~+~ 2.424,
\label{eq:Morgan07}
\end{equation}

\begin{equation}
M_V = 1.061 ~-~ 0.961~P ~-~ 0.044~\phi^{(s)}_{21} ~-~ 4.447~A_4.
\label{eq:K98}	
\end{equation}

\noindent 

For convenience, one can transform the coefficients from cosine series phases into sine series using the following relation: 
\begin{equation}
\phi^{(s)}_{jk} = \phi^{(c)}_{jk} - (j - k) \frac{\pi}{2}.
\label{eq:K98}	
\end{equation}

\begin{table*}
\caption{Fourier coefficients $A_{k}$ for $k=0,1,2,3,4$, and phases $\phi_{21}$, $\phi_{31}$ and $\phi_{41}$, for RRab and RRc stars. The numbers in parentheses indicate the uncertainty on the last decimal place. Also listed is the deviation parameter $D_{\mbox{\scriptsize m}}$ for V8 (see  \citet{Jurcsik1996}).}
\centering                   
\begin{tabular}{lllllllllr}
\hline
Variable ID     & $A_{0}$    & $A_{1}$   & $A_{2}$   & $A_{3}$   & $A_{4}$   &
$\phi_{21}$ & $\phi_{31}$ & $\phi_{41}$ 
&  $D_{\mbox{\scriptsize m}}$ \\
     & ($V$ mag)  & ($V$ mag)  &  ($V$ mag) & ($V$ mag)& ($V$ mag) & & & & \\
\hline
       &       &   &   &   & RRab star    & &        
   &             &       \\
\hline

V8 &14.843(1) &0.295(1) &0.145(2) &0.093(1) &0.039(1) &4.348(1) &8.846(2) &7.0565(4) &2.1  \\

\hline
             &            &           &           &           & RRc stars &           
 &             &             &\\
\hline
V5  &14.790(1) &0.240(1) &0.010(1) &0.024(1) &0.014(1) &5.073(82) &4.638(35) &2.693(59) \\

V7  &14.897(1) &0.156(1) &0.018(1) &0.004(1) &0.002(1) &4.761(34) &3.278(146)& 1.806(318) \\

V9  &14.819(1) &0.244(1) &0.017(1) &0.017(1) &0.013(1) &4.921(64) &4.532(65)& 3.072(83)\\

V25 &14.622(1) &0.186(1) &0.006(1) &0.009(1) &0.006(1) &3.755(177) &4.456(110) & 2.463(176) \\

V34  &14.806(1) &0.173(1) &0.009(1) &0.016(1) &0.010(1) &6.382(108 )&4.781(63) &2.748(93) \\

V35 &14.807(1) &0.088(1) &0.006(1) &0.002(1) &0.002(1) &5.153(121) &3.344(380) &2.547(389) \\

V54 &14.897(1) &0.067(1) &0.010(1) &0.001(1) &0.001(1) &4.323(78) &5.574(5805) &6.405(5926) \\

\hline	
\hline
\end{tabular}
\label{tab:fourier_coeffs}
\end{table*}

\begin{table*}
\footnotesize
\begin{center}
\caption[] {\small Physical parameters obtained from the Fourier fit for the RRab and RRc stars. The numbers in
parentheses indicate the uncertainty on the last decimal place. See $\S$ \ref{decomposition} for a detailed discussion.}
\label{fisicos}
\hspace{0.01cm}
 \begin{tabular}{llllllll}
\hline 
 &  &  & RRab star &  & & \\
\hline
Star&[Fe/H]$_{\rm ZW}$&[Fe/H]$_{\rm UVES}$ & $M_V$ & log~$T_{\rm eff}$  & log$(L/{L_{\odot}})$ & $M/{M_{\odot}}$&$R/{R_{\odot}}$\\

\hline
V8$^1$  &-1.603 $\pm$ 0.002 &-1.536 $\pm$ 0.002& 0.378 $\pm$ 0.016& 3.794 $\pm$ 0.008 & 1.749 $\pm$ 0.006 & 0.68 $\pm$ 0.07 & 6.50 $\pm$ 0.05  \\
\hline

 &  &  & RRc stars &  & & \\
 \hline
Star&[Fe/H]$_{\rm ZW}$&[Fe/H]$_{\rm UVES}$ & $M_V$ & log~$T_{\rm eff}$  & log$(L/{L_{\odot}})$ & $M/{ M_{\odot}}$&$R/{R_{\odot}}$\\

\hline
V7                 &-1.53(27) &-1.44(30) &0.611(5) &3.866(1) &1.655(2) &0.524(6)  &4.180(9)   \\
V25                &-1.81(15) &-1.81(19) &0.542(6) &3.854(1) &1.683(3) &0.412(3)  &4.571(13)  \\
V35                &-1.39(48) &-1.28(48) &0.586(6) &3.867(1) &1.666(3) &0.517(9) &4.215(12)  \\
V54$^2$            & - & - &0.652(6) & 3.883(33)&1.639(2) & 0.449(172)& 3.804(10) \\

\hline
Weighted Mean&-1.72&-1.66&0.586&3.856& 1.665&0.45 & 4.28\\
$\sigma$ &$\pm$ 0.12  &$\pm$ 0.15 & $\pm$ 0.003 &$\pm$ 0.001 &$\pm$ 0.001& $\pm$ 0.01 & $\pm$ 0.01 \\
\hline
\end{tabular}
\end{center}
\raggedright
\center{
1. The uncertainties in the parameters come from the uncertainties of the calibrations themselves.\\
2. Not included in the averages of the physical parameters.\\
}
\end{table*}

For comparison, we have transformed [Fe/H]$_{\rm ZW}$
on the \cite{Zinn1984} metallicity scale into the UVES scale
using the equation [Fe/H]$_{\rm UVES}$= $-0.413$ + 0.130~[Fe/H]$_{\rm
ZW} - 0.356$~[Fe/H]$_{\rm ZW}^2$ \citep{Carretta2009}. \\

The values of $M_{V}$ reported in Table \ref{fisicos} have been transformed to luminosities using the following equation:

\begin{equation}
    \rm log(L/L_{\odot}) = -0.4(M_{V} - M^{\odot}_{bol} + BC).
\end{equation}

To calculate the bolometric correction, we made use of the formula $BC = 0.06[\rm Fe/H]_{ZW} + 0.06$ reported in \citet{SanCac1990}. We have adopted the value of  $M^{\odot}_{bol}$ = 4.75 mag.

To estimate the effective temperature of the RRab stars we employed the calibration given by \citet{Jurcsik1998}:

\begin{equation}
    \rm log(T_{\rm eff}) = 3.9291-0.1112(V-K)_{0}-0.0032[\rm Fe/H],
\end{equation}

\noindent
where 

\begin{equation}
(V-K)_{0} = 1.585+1.257P-0.273A_{1} - 0.234\phi_{31}^{(s)} + 0.062\phi_{41}^{(s)}. 
\end{equation}

For the RRc stars, the calibration of \citet{Simon1993} was used:

\begin{equation}
    \rm log(T_{\rm eff}) = 3.7746 - 0.1452\rm log(P) + 0.0056\phi_{31}^{(c)}.
\end{equation}
Once we have estimated the effective temperature, the period and the luminosity of the RR Lyrae stars, we can also estimate their masses using: 
$\rm log(M/M_{\odot}) = 16.907 - 1.47\rm logP_{F} + 1.24\rm log(L/L_{\odot}) - 5.12\rm log(T_{\rm eff})$
as given by \citet{vAlbada1971} where logP$\rm_{F}$ is the fundamental period, and their radii with $L  = 4\pi R^{2} \sigma T^{4}$. These values are also reported in Table \ref{fisicos}.

Stars with \emph{Bl} designation in Table \ref{variables}, are Blazhko variables. We report their Fourier coefficients and physical parameters but they were not taken into account in the physical parameters calculations. Since the Fourier decomposition for V54   yielded an  anomalous value for $\phi_{31}$,
on which the metallicity, temperature and hence the mass and radius depend, we decided not to include this star in the average of these parameters. However, we included it in the calculation of the average distance, since it does not depend on $\phi_{31}$. The resulting physical parameters are summarised in Table \ref{fisicos}. \\

\subsection{An alternate formulation for [Fe/H]}

We performed our calculations of [Fe/H] via the Fourier parameters using eqs. \ref{eq:JK96} and \ref{eq:Morgan07} for the RRab and RRc stars respectively, mainly for the sake of homogeneity, since these formulations have been systematically used by our group for a large number of clusters (e.g. \cite{Are2017}). However, new formulations have been proposed by \citet{Nemec2013}, in their eqs. 2 and 4 for the RRab and RRc respectively.
We have used these new formulations to calculate [Fe/H]$_{\rm ZW}$ as -1.573 for the only RRab star and -1.74 for the three RRc stars. These values are, within the corresponding  uncertainties of both formulations, in very good agreement with the averages reported in Table \ref{fisicos}. \citet{Nemec2013} have noted the good agreement between the two formulations for clusters with [Fe/H] < -1.0, as it seems to be the case for M13, and that larger discrepancies are to be found for decreasing metallicities. For [Fe/H]$\sim -2.0$ the differences can be as large as $\sim$0.3 dex.

\begin{figure} 
\includegraphics[width=8.0cm,height=4.3cm]{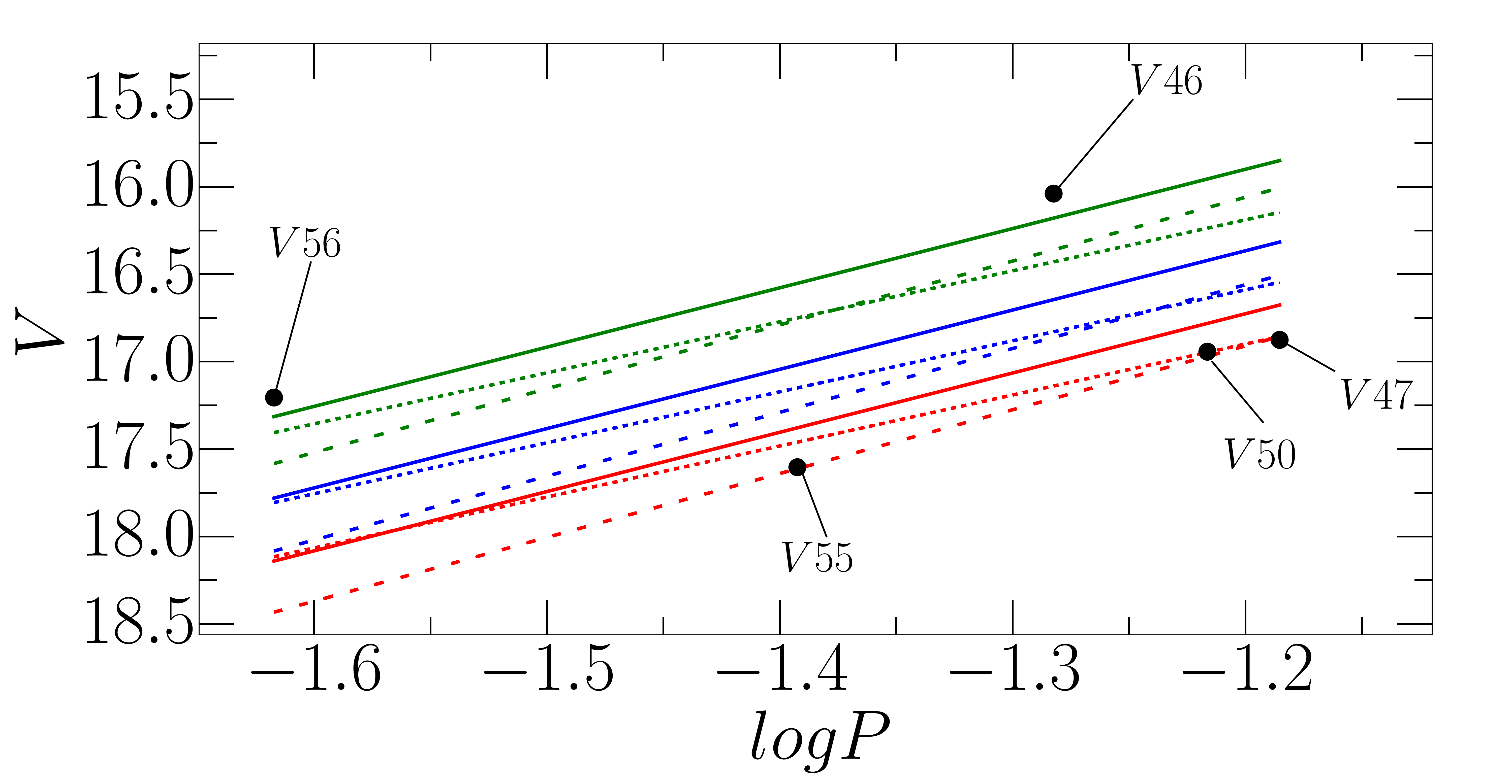}
\caption{Period-Luminosity relation in $V$ for SX Phe stars in M13. Coloured solid lines correspond to the P-L relation derived by \protect\citet{CohenSara2012}, the long-dashed lines to the relation derived by \protect\citet{Poretti2008} and the short-dashed lines correspond to the P-L relation derived by \protect\cite{Arellano2011}. The colours red, blue and green correspond to the fundamental, first overtone and second overtone respectively.}
\label{vlogp}
\end{figure}

\section{On the metallicity of M13}

M13 is a metal-poor cluster of the OoII type. From the Fourier decomposition of the RR Lyrae light curves, we found an overall average of  <[Fe/H]$_{\rm ZW}$> = $-1.58 \pm 0.09$. Other values for the metallicity found in the literature include: --1.65 \citep{Zinn85}, --1.65 \citep{Djor1993}, --1.63 \citep{Marin2009}, --1.58 \citep{Carretta2009} and --1.53 \citep{Harris1996}, which are in good agreement with our own results.

\section{On the distance to M13}

The distance to M13 has been a matter of controversy. In the literature, we have found an assortment of values that range from 7.1 kpc \citep{Harris1996}, 7.4 kpc (\citealt{Paust2010}, \citealt{Vandenberg2013}), 7.7 kpc \citep{Hessels2007} to 8.0 kpc \citep{Violat2005}. In the following, we explain our own results for the determination of the distance to M13 by using several independent distance indicators. 

\subsection{From the RR Lyrae stars}
We started by adopting the reddening value reported by \citet{Harris1996} of $E(B-V)$ = 0.02. With this value, and using the independent calibrations for RRab and RRc stars to calculate $Mv$ mentioned in $\S$ \ref{decomposition}, we obtained the distances $d$ = 7.6 kpc for the only known RRab in the cluster and $<d>$ = 6.8 $\pm$ 0.3 kpc for the RRc stars. We have also made use of the P-L relation for RR Lyrae stars in the $I$ filter \citep{Catelan2004} $M_I = 0.471-1.132~ {\rm log}~P +0.205~ {\rm log}~Z$, with ${\rm log}~Z = [M/H]-1.765$; $[M/H] = \rm{[Fe/H]} - \rm {log} (0.638~f + 0.362)$ and log~f = [$\alpha$/Fe] \citep{Sal93}, from which we derived a distance of 7.0 $\pm$ 0.2 kpc. When taking these approaches, we made use of all the RR Lyrae stars with the exception of V31 and V36 since a double-mode effect appears to be present in their light curves. 

\subsection{From the Type II Cepheids}
There are three known Type II Cepheids in M13, namely V1, V2 and V6. We made use of Fourier decomposition of V1 and V2 in order to estimate an intensity weighted mean for the apparent magnitude. Since the light curve of V6 is incomplete, we adopted the value provided by \citet{Kopacki2003} of $V$ = 14.078. We used the P-L relation for these type of stars $M_V = -1.64(\pm0.05)$ $\rm logP$ + 0.05($\pm$0.05) as derived by \citet{Pritzl2003}, from where we found a distance of 7.1 $\pm$ 0.6 kpc. 

\subsection{From the SX Phe stars}
\label{sxphe}
In Fig. \ref{vlogp} we show the P-L relation for the SX Phe stars in the log $P$-$V$ plane. There are four known SX Phe stars in M13 listed in the catalogue of variable stars in globular clusters \citep{Clement2001}, plus two that are newly reported in this work (V55 and V56). Three versions of the P-L relation are shown in the figure. This relation is helpful for confirming star membership in the cluster as well as for identifying pulsation modes, and hence to estimate the mean distances to qualified cluster members. Making use of the P-L relation derived by \citet{Poretti2008}; $M_V$ = -3.65($\pm$0.07)~ log~$P$-1.83($\pm$0.08) we found $<d>$ = 7.2 $\pm$ 0.7 kpc.
From this diagram we conjecture that V47, V50 and V55 are pulsating in the fundamental mode while V46 and V56 are likely second overtone pulsators.

\subsection{From the binary V57}

Via the model fitting for the contact binary V57, a distance of 6.9 $\pm$ 0.2 kpc was estimated (see $\S$ \ref{V57}). 

\subsection{Distance from the ZAHB fitting}

According to \citet{Vandenberg2013}, a proper fitting of the theoretical Zero Age Horizontal Branch (ZAHB) and the Turn Off (TO) to the observed stellar distributions, gives a reliable apparent magnitude modulus. The apparent modulus that we found in this way for M13 was $\mu$ = 14.34 corresponding to a distance $d$ = 7.1 kpc. The average of the distance values obtained by the aforementioned methods, listed in Table \ref{distance}, is $<d>$ =7.1 $\pm$ 0.1 kpc.

\subsection{Luminous Red Giants as distance indicators}

The luminosity of the brightest stars at the TRGB can in principle be used to estimate the distance to a stellar system. This method was originally envisaged to determine distances to nearby galaxies \citep{Lee1993}.  As calibrated by \citet{SalCas1997}, the bolometric magnitude of the tip of the RGB is:

\begin{equation}
\label{TRGB}
M_{bol}^{tip} = -3.949\, -0.178\, [M/H] + 0.008\, [M/H]^2,
\end{equation}
\\

\noindent
where $[M/H] = \rm{[Fe/H]} - \rm {log} (0.638~f + 0.362)$ and log~f = [$\alpha$/Fe]
\citep{Sal93}.

 The  method is very sensitive to the selection of the stars to be used, and on the other hand, it has to be considered that the brightest stars in a given cluster may not be at the very tip of the RGB in the CMD, but rather lower by a certain magnitude. \cite{Viaux2013}  argued, on the grounds of helium ignition delay in low-mass stars, and the resulting extension of the red giant branch, that for the case of M5 the brightest stars are between 0.04 and 0.16 mag below the TRGB. The suggested offset was confirmed by the non-canonical models of \cite{Arceo2015}, whom from the analysis of 25 globular clusters concluded that the theoretical TRGB is in average about 0.26$\pm$0.24 bolometric magnitudes brighter than the one observed.
 
  Using only the two brightest stars near the tip of the RGB in the CMD (V11 and V42), and making use of eq. \ref{TRGB} without further assumptions, one finds a distance of 8.1 kpc, which is too high relative to the distance found by all methods previously discussed. We figured that, to reproduce the mean distance in Table \ref{distance}, 7.1 kpc, we need to assume that the true TRGB is some 0.25 magnitudes brighter than V11 and V42, in agreement with the predictions of \cite{Arceo2015}. However, using the new calibration of \citet{Mould2019}, whom from $Gaia$-DR2 data calculated that the TRGB is close to $M_I \sim -0.4$ mag,
  we found that in order to get a distance of 7.1 kpc for V11 and V42, we need a correction of only 0.08 mag, in good agreement with the calculations of  \cite{Viaux2013}.   \\

\begin{table}
\begin{center}

\caption{Distance comparison to M13 from the different methods used in this work.}
\label{distance}

\begin{tabular}{lc}
\hline
Method & Distance [kpc]        \\
\hline
RRab Fourier decomposition      & 7.6$^1$ \\
RRc  Fourier decomposition      & 6.8 $\pm$ 0.3 \\        
RRab / RRc $I$-band P-L  & 7.0 $\pm$ 0.2 \\        
SX Phe P-L                          & 7.2 $\pm$ 0.7 \\        
Type II Cepheids P-L               & 7.1 $\pm$ 0.6 \\
Level of the ZAHB      & 7.1 $\pm$ 0.1 \\
Eclipsing Binary V57      & 6.9 $\pm$ 0.2 \\
\hline
\bf{Weighted mean} & 7.1 $\pm$ 0.1 \\  

\hline
\end{tabular}
\end{center}
\raggedright
\center{\quad 
1. Based only on one RRab. Not included in the average.}

\end{table}

\begin{figure*} 
\includegraphics[width=16.0cm,height=4.3cm]{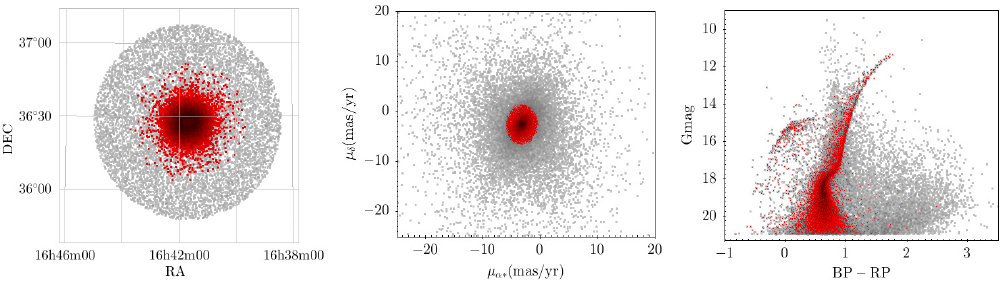}
\caption{ The red dots correspond to member stars and the gray dots to field stars according to the astrometric membership determination. Left panel: cluster field; central panel: vector point diagram (VPD) of proper motions; right panel: color-magnitude diagram (\textit{Gaia} photometric system).}
\label{M13_GAIA}
\end{figure*}

\section{Star membership using \textit{Gaia}}
\label{gaia_members}

\begin{figure} 
\includegraphics[width=8.0cm,height=7.5cm]{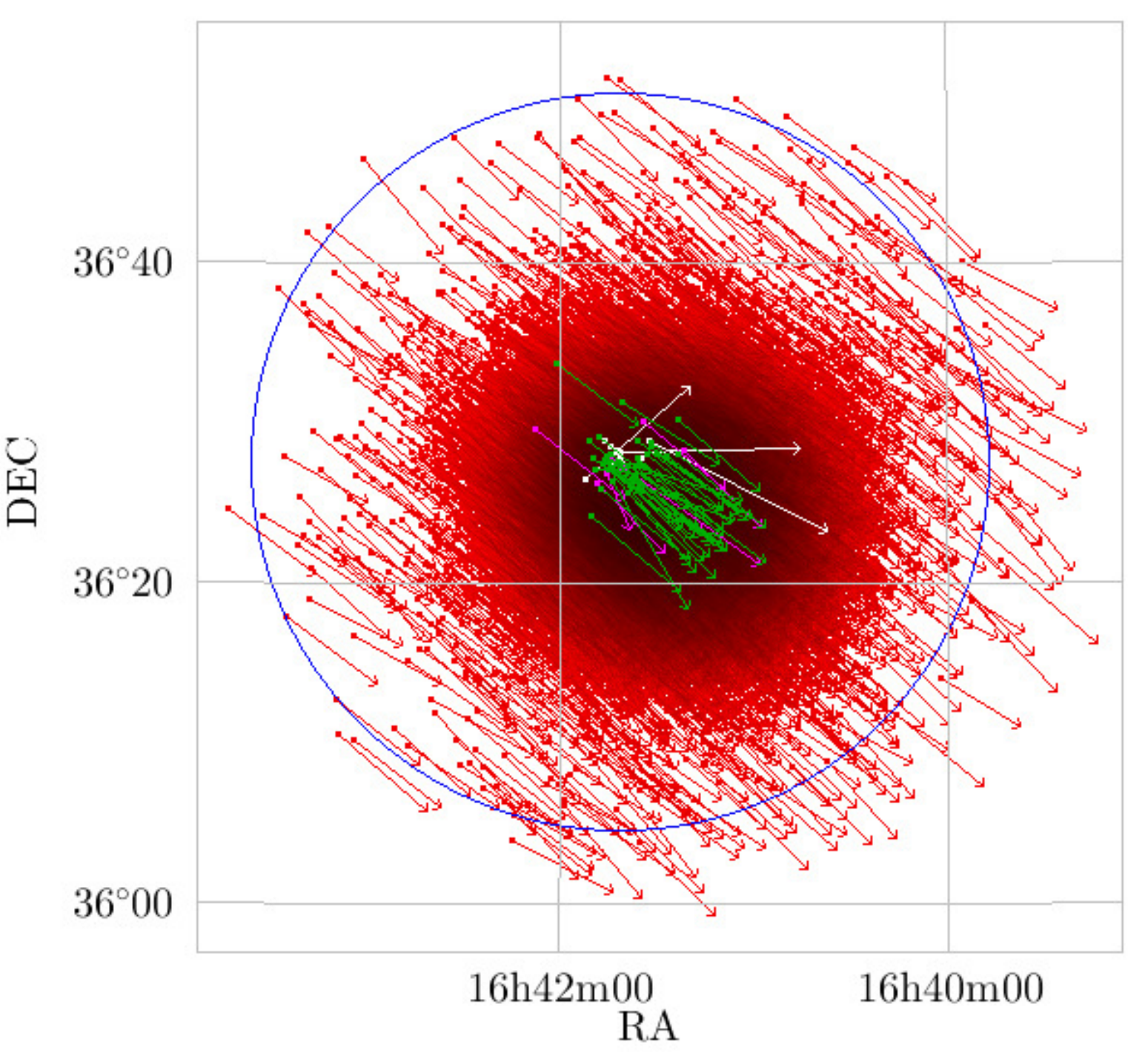}
\caption{Projection on the sky of the proper motion vectors. Red arrows: cluster members; green arrows: known  variable stars,  pink arrows are newly discovered variables. Vectors have been enlarged 100000x for visualisation purposes.}
\label{VAR_MPs}
\end{figure}

 In this section to ascertain cluster membership we used an approach based upon the high quality astrometric data available in \textit{Gaia}-DR2 \citep{Gaia2018}. The method is based on the  Balanced Iterative Reducing and Clustering using Hierarchies (BIRCH)  algorithm \citep{Zhang1996} in a four-dimensional space of physical parameters -positions and proper motions- that detects groups of stars in that 4D space. A 4D gaussian ellipsoid is fitted for each group, stars outside 3$\sigma$ are rejected. In order to decide if the remaining stars are members of the cluster, their positions are plotted in celestial coordinates, in the vector point diagram (VPD) of the proper motions and in the CMD (of \textit{Gaia} photometric system). If all three plots are consistent with those of a globular cluster, the stars are considered members of the cluster (Fig. \ref{M13_GAIA} and Fig. \ref{VAR_MPs}). More details about the method and some applications on other clusters  will be published elsewhere by Bustos Fierro \& Calder\'on. We applied this method on a field of 40 arcmin of radius around the center of M13 containing 52,802 stars. We found 23,070 stars most likely members of the cluster. In Fig. \ref{M13_GAIA} we display the result of this membership determination.  

We then cross-matched the member stars with our collection of stars with light curves in the FoV of our M13 images. We found 7,630 stars in common that were used to build the clean CMD's of Fig. \ref{CMD_6205} (central panel). 

We also cross-matched the member stars with the variables reported by \citet{Clement2001} and the new variables discovered in this work in order to confirm the membership status of all known variables in M13. Fig. \ref{VAR_MPs} plots the positions and proper motions of all members and the
variables.

\section{The CMD of M13}
\label{CMD}
\begin{figure*}
\begin{center}
\includegraphics[width=\textwidth, height=9cm]{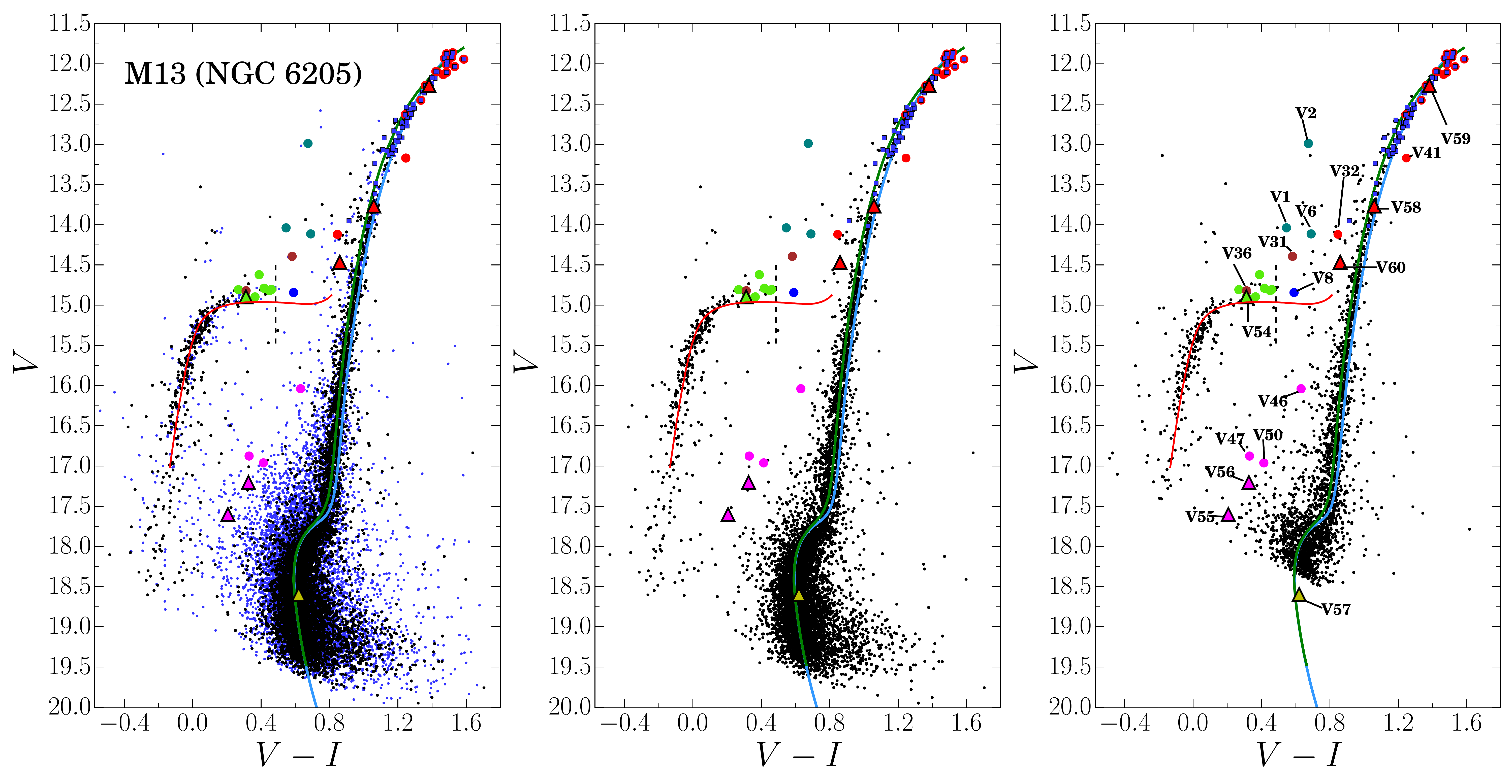}
\caption{Colour-Magnitude Diagrams of M13.  The left panel shows the CMD with all the stars in the FoV of the IAC images. Black and blue dots represent the member and non-member stars (see $\S$ \ref{gaia_members}). The central panel shows only those stars with a high probability of being members of the cluster. The right panel shows the member stars as measured from Hanle data exclusively. The colour markers are the same in all three panels. The blue circle corresponds to the only known RRab star in the cluster. Green and magenta circles correspond to RRc and SX Phe stars respectively. Red circles represent semi-regular variables. Brown circles correspond to RRd stars and the teal circles to Type II Cepheid stars. The triangle markers correspond to newly discovered RRc (V54), SX Phe (V55 and V56), W Uma (V57) and SR (V58, V59 and V60) stars. The blue squares near the top of the RGB are 70 stars whose membership has been determined by means of radial velocity from \textit{Gaia}. The light blue isochrone is from \citet{Vandenberg2014}, with  Y = 0.25 and [$\alpha$/H] = 0.4, corresponding to an age of 12.6 Gyrs. The green isochrone is based on the models of \citet{Marigo2017}
and corresponds also to an age of 12.6 Gyrs and to a <[Fe/H]>$_{RR}$ = -1.58. The isochrones and ZAHB have been shifted to a distance of 7.1 kpc (see Table \ref{distance}) and reddened by $E(B-V)=0.02$, althought the ZAHB (red line) needed and extra 0.02 for a better fit. The dashed vertical line represents the red edge of the first overtone instability strip. See $\S$ \ref{CMD} for a discussion.} 

\label{CMD_6205}
\end{center}
\end{figure*}

The Colour-Magnitude Diagrams (CMD) in the mosaic of Fig. \ref{CMD_6205} illustrate the cleaning process of field stars. The left panel shows all stars in the field of the cluster with the photometric measurements in this work. Black symbols are used for member stars identified as described in $\S$ \ref{gaia_members}.  The central and right panels illustrate the distribution of member stars, for the IAC and Hanle data respectively. The distribution of variable stars and the isochrone and ZAHB fitting to the stellar distributions, assuming the parameters derived in this paper, are also shown. Note that the positions of the variable stars are the same in all the CMDs since their mean intensity weights were calculated using both sets of data (IAC and Hanle). 

\subsection{The structure of the horizontal branch of M13}

The HB of M13 displays mostly a prominent blue tail and the redder components seem evolved and lie well above a ZAHB. The RR Lyrae population is dominated by first overtone RRc pulsators. Only one RRab is known and two double-mode or RRd stars are also present. The distribution of RRab-RRc stars shows a clear segregation around the first overtone red edge, represented as a vertical dashed line identified in several other clusters, which seems to be the rule among OoII clusters \citep{Arellano2018}. A parameter that helps to describe the morphology of the HB is the Lee index \citep{Lee1990} defined as $\mathcal{L}$ = (B-R)/(B+V+R), where B and R are the number of stars to the blue or red sides of the instability strip respectively,
and V is the number of variable stars within the instability strip. In the case of M13 we found this value to be $\mathcal{L}$ = 0.95, which is consistent with a OoII type cluster with the two pulsation modes well split.

\section{Summary and Conclusions}

In this work we have obtained high-precision photometry of over 16,000 point sources in the field of the globular cluster M13, in our $V$ and $I$-band reference images. We have been able to retrieve most of the known variables cited in the literature. Of particular interest is the RR Lyrae star population since the Fourier decomposition of their light curves allowed us to determine  the metallicity of M13 by two independent methods, from which we derived an average of <[Fe/H]$_{\rm ZW}$> = $-1.58 \pm 0.09$. 

From our two-colour photometry we constructed a clean CMD based on 
the recent $Gaia$-DR2 \citep{Gaia2018} proper motions and radial velocities, on
which we overlaid two isochrones that correspond to an age of 12.6 Gyrs and a metallicity of <[Fe/H]>$_{RR}$ = -1.65 and a theoretical ZAHB with the corresponding metallicity. The isochrones come from different sets of model atmospheres and yet they are practically indistinguishable from one another. 

We discovered 7 new variables: one RRc (V54), two SX Phe (V55 and V56), one contact binary (V57), and three SR (V58, V59 and V60) star. From the orbital solution of the contact binary, we were able to derive the masses, radii and effective temperatures of the primary and secondary components. We also found 15 stars which seem to display mid- to long-term variations but that have been retained as candidates waiting for confirmation by more appropriate data. Based on its high radial velocity, we identified a runaway star in the field of M13 which is discussed in Appendix \ref{RAstar}. This star appears to be passing by and and overtaking the cluster.

We report the double-mode nature of the star V31 and confirm the double-mode nature of V36. In our analysis of V36, we found two frequencies that are consistent with the ones found by \citet{Kopacki2003}. In the case of V31, we also found two distinct periods and at least one seems to be non-radial. 

From the P-L relation for the SX Phe stars, we found that is likely that V47, V50 and V55 are pulsating in the fundamental mode and V46 and V56 in the second overtone. 

The position of the only RRab in M13 (V8) on the Bailey diagram and on the CMD above the HB, suggests that this star is in an advanced evolutionary stage moving towards the AGB. 
The period of V8, the large positive value of $\mathcal{L}$ (long blue tail of the HB), and the clear segregation between RRab and RRc stars on the HB, confirm the nature of M13 as an OoII type cluster. 

By using seven independent methods such as the $M_V$ values obtained from the Fourier decomposition of RRab and RRc stars, the P-L relation for RR Lyrae stars in the $I$ filter, the P-L relation for the SX Phe, the P-L relation for the CW stars, the position of the theoretical ZAHB and the orbital solution of the contact binary V57, we were able to estimate the mean distance to M13, d = 7.1 $\pm$ 0.1 kpc.

From $Gaia$-DR2 proper motion data we found 23,070 stars that are likely cluster members. We also re-confirm the membership status of all variables referred in this work and we give the $Gaia$-DR2 identifier for all of them. From the analysis of the eighteen stars of the catalog of \citet{Clementini2018} that are in the field (see Table 4) we conclude that four of them were previously known (namely V5, V7, V8 y V9), seven of them do not show variability, six of them are fainter than the limit of our observations and one is out of the FoV of our frames.

\section*{Acknowledgements}

We are grateful to Prof. Don VandenBerg for his valuable insights on the ZAHB and isochrone models. We thank the staff of IAO, Hanle and CREST, Hosakote, that made these observations possible. The facilities at IAO and CREST are operated by
the Indian Institute of Astrophysics, Bangalore.
This project was partially supported by DGAPA-UNAM (Mexico) via grant IN106615-17.
DD thanks CONACyT for the PhD scholarship. We have made extensive use of the SIMBAD, ADS services, and of "Aladin sky atlas" developed at CDS, Strasbourg Observatory, France.
\citep{Bonnarel2000} and TOPCAT \citep{Taylor2005}.




\bibliographystyle{mnras}
\bibliography{6205_biblio} 



\appendix
\label{appendix}

\section{A runaway star in the field of M13}
\label{RAstar}

\begin{figure} 
\includegraphics[width=8.0cm,height=6.0cm]{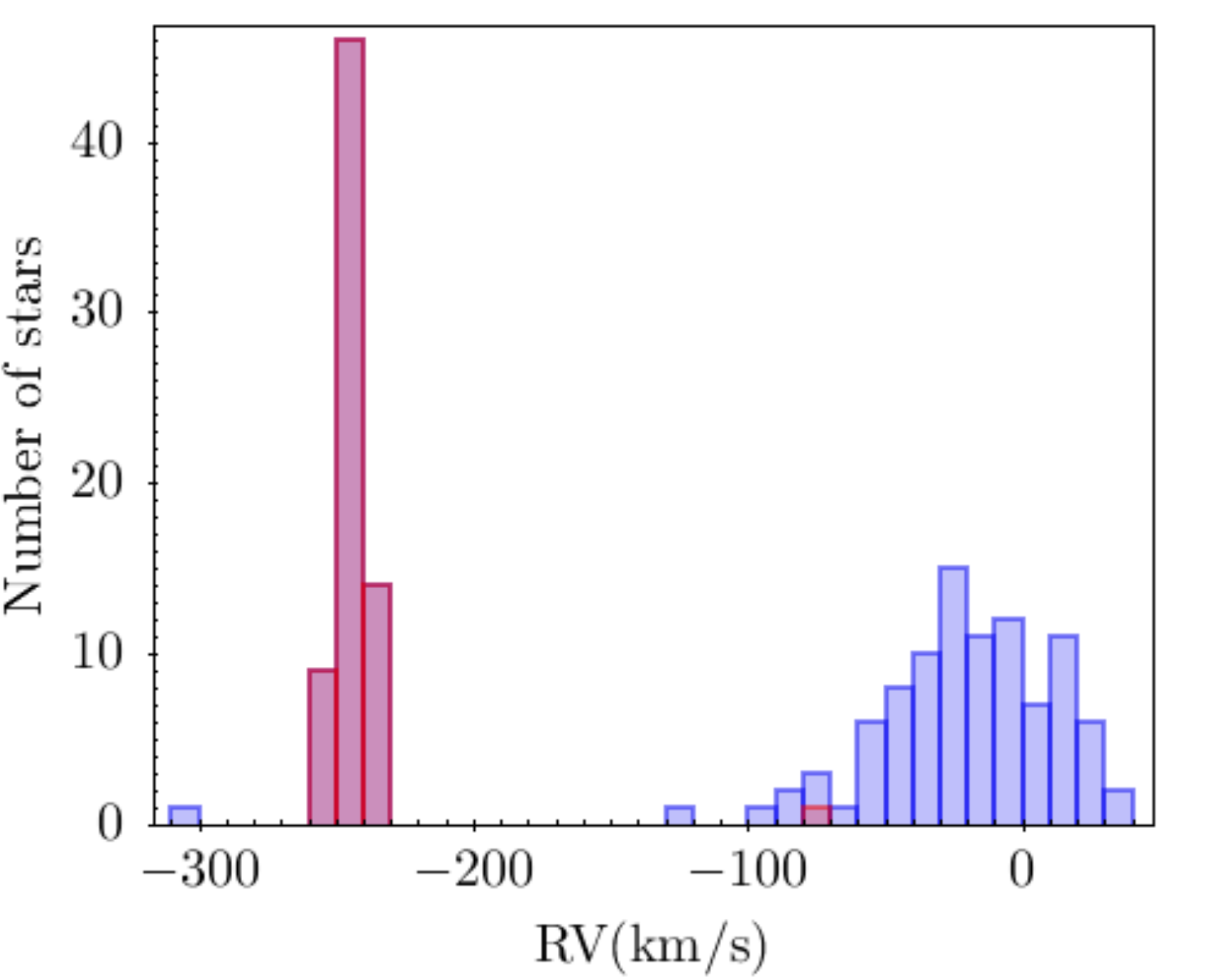}
\caption{Histogram of radial velocities for 166 stars measured with \textit{Gaia} in the FoV of M13. Purple bars at $V_r \sim -250$ km/s are cluster members. Light blue bars are non-members. }
\label{M13_VR}
\end{figure}

\begin{figure} 
\includegraphics[width=8.0cm,height=6.0cm]{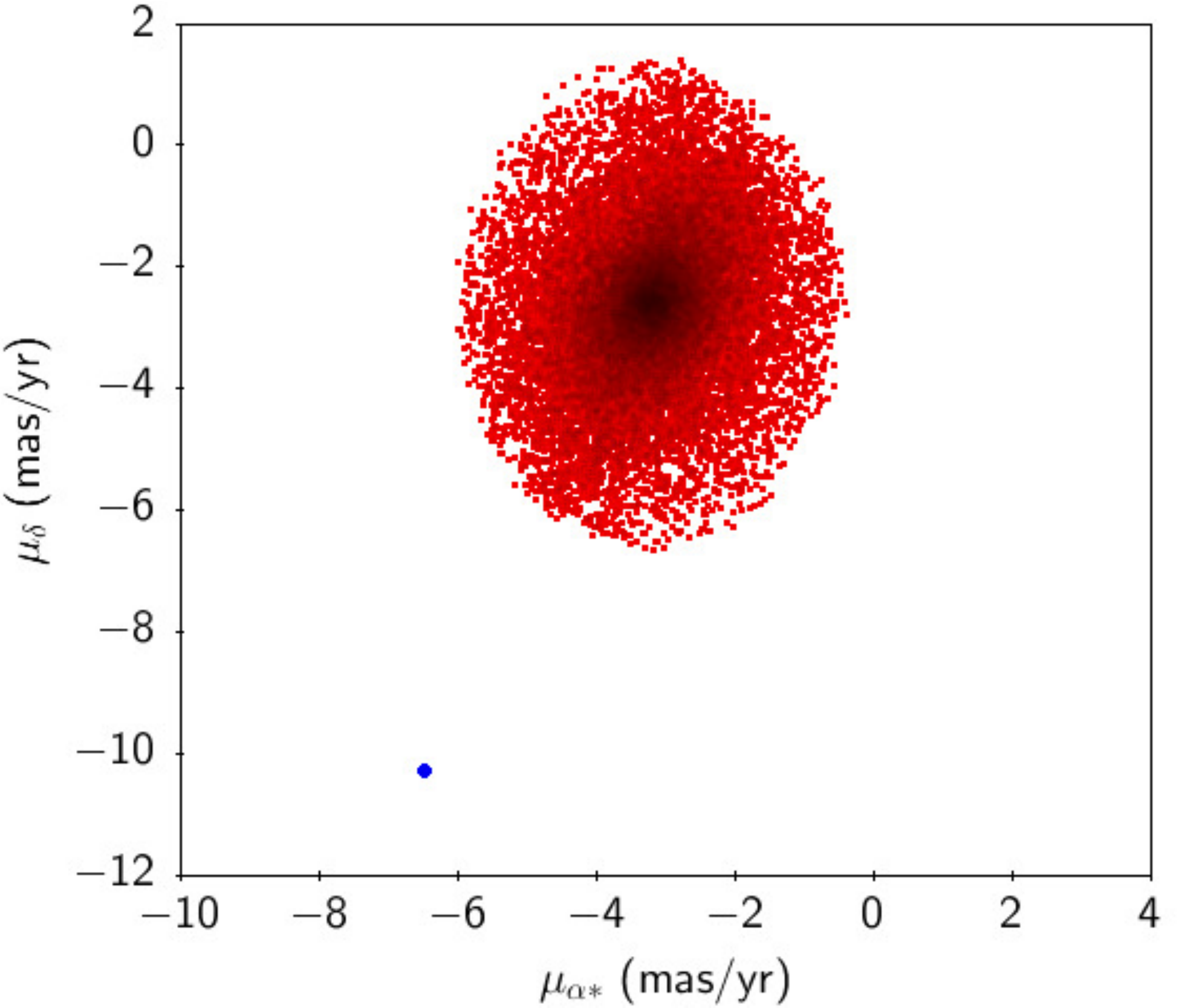}
\caption{VPD of the members of M13. The blue dot shows the proper motion of the HRVS. }
\label{RAwayMP}
\end{figure}

After the membership determination, we analysed the radial velocities distribution in the field of M13. There are 166 stars with radial velocity measured in \textit{Gaia}-DR2, 70 of which were identified as members of the cluster. The radial velocity histogram in Fig. \ref{M13_VR} shows a coherent concentration of the cluster members at $V_r \sim -250$ km/s whereas the non-members are neatly separated at lower velocities. In this figure, there is one noticeable High Radial Velocity Star (HRVS) at $V_r \sim -310$ km/s. This HRVS is the \textit{Gaia}-DR2 source 1328140781417544832, RA (ICRS) 250.97068o, DE (ICRS) 36.48989o, $\mu_\alpha$=(-6.468 $\pm$ 0.029)mas/yr,  $\mu_\delta$=(-10.296$\pm$0.035)mas/yr  $V_r$=(-304.74$\pm$1.67)km/s, $\varpi$=(0.1708$\pm$0.0198)mas (Fig. \ref{RAwayMP}).

\begin{figure} 
\includegraphics[width=8.0cm,height=7.0cm]{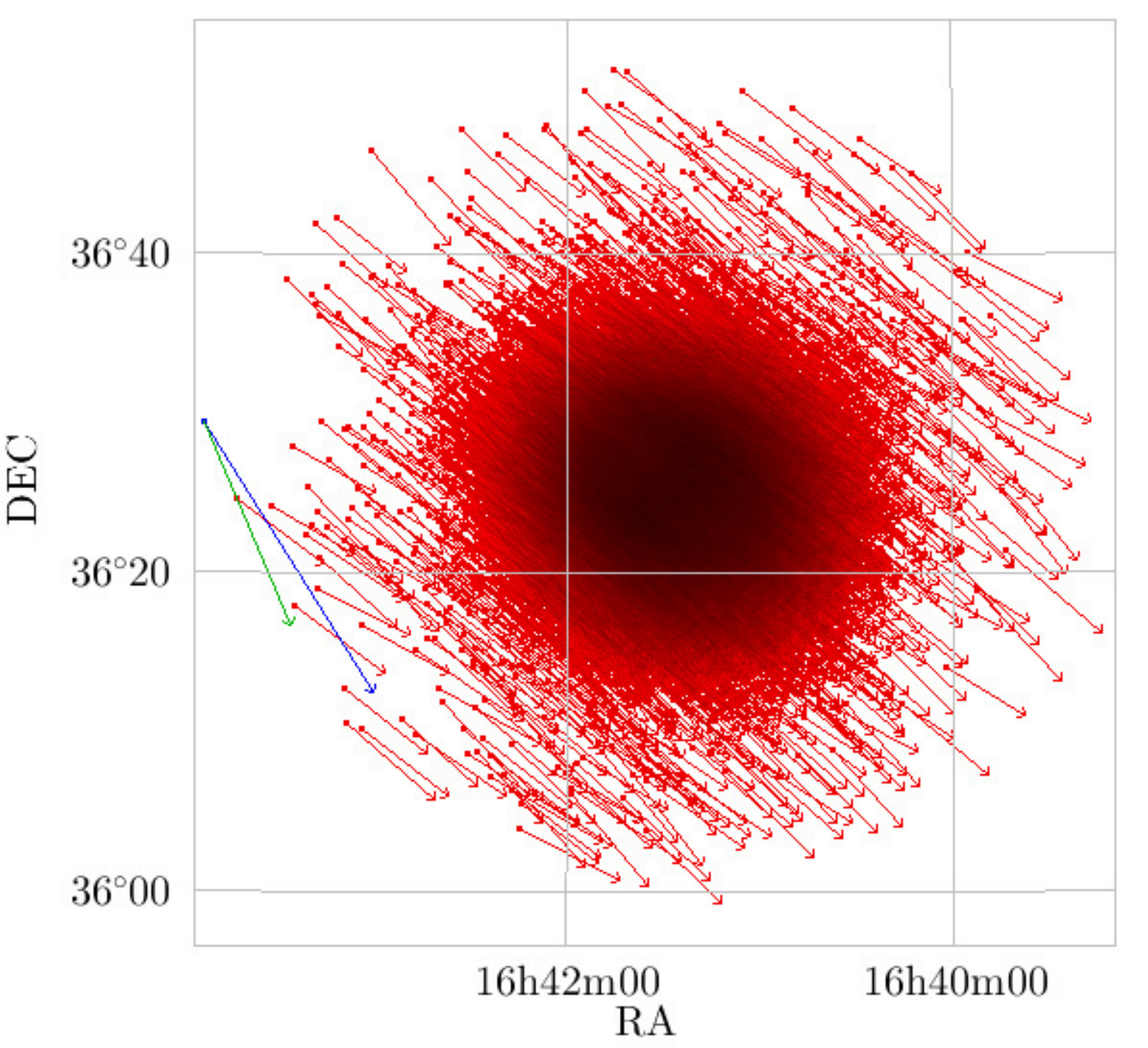}
\caption{Projection on the sky of the proper motion vectors. Red arrows: cluster members; blue arrow: HRVS; green arrow: HRVS relative to the mean proper motion of the cluster. Vectors have been enlarged 100000x for visualisation purposes.}
\label{RAway}
\end{figure}

In Fig. \ref{RAway} we display the positions and proper motion vectors for the cluster members and the HRVS. If it was a member star escaping the cluster, it is expected to have a relative proper motion approximately radial from the centre of the cluster, but this is not the case. Instead, all this evidence suggests that the HRVS is passing by very close to the cluster and overtaking it in a rather tangential way. It is also noticeable a star among the members with discrepant radial velocity (MDRV), it corresponds to the \textit{Gaia}-DR2 source 1328033613402105728, RA (ICRS) 250.33156o, DEC (ICRS) 36.35434o, $\mu \alpha$*=(-4.732$\pm$0.039)mas/yr, $\mu \delta$=(--1.713$\pm$0.054)mas/yr, RV=(--73.02$\pm$1.67)km/s,  $\varpi$=(0.2854$\pm$0.0234)mas. If this star is indeed a cluster member, we are unable to figure out the reason for the discrepancy with the available information, if not it could be a field star that contaminates the
sample of member stars.

\bsp	
\label{lastpage}
\end{document}